\documentclass[iop, numberedappendix]{mn2e} 

\usepackage[pdftex]{graphicx}

\usepackage{amsmath, amsthm, amssymb}
\let\bibhang\relax
\usepackage{natbib}
\usepackage{subfigure}
\usepackage{upgreek}
\usepackage{accents}
\usepackage{color}
\usepackage{hyperref}
\usepackage{dblfloatfix}
\usepackage[T1]{fontenc}
\usepackage{aecompl}
\usepackage{enumitem}

\hypersetup{
  colorlinks   = true, 
  urlcolor     = blue, 
  linkcolor    = blue, 
  citecolor   = blue 
}

\newcommand*{\dt}[1]{%
  \accentset{\mbox{\large\bfseries .}}{#1}}

\newcommand{\ares}{\textsc{ares}}

\newcommand{\dpl}{\texttt{dpl}}

\newcommand{\meanslope}{\langle \delta T_b^{\prime} \rangle}
\newcommand{\meanslopelo}{\langle \delta T_b^{\prime} \rangle_{\mathrm{lo}}}
\newcommand{\meanslopehi}{\langle \delta T_b^{\prime} \rangle_{\mathrm{hi}}}
\newcommand{\hwhmdiff}{\mathcal{A}}
\newcommand{\squash}{\mathcal{W}}

\newcommand{\two}{\textsc{ii}}
\newcommand{\three}{\textsc{iii}}

\newcommand{\Lya}{\text{Ly-}\alpha}
\newcommand{\Lyn}{\text{Ly-}n}

\newcommand{\NHI}{N_{\text{H } \textsc{i}}}

\newcommand{\Jlw}{J_{\text{LW}}}

\newcommand{\SFR}{\dot{M}_{\ast}}
\newcommand{\SFRII}{\dot{M}_{\ast,\two}}
\newcommand{\SFRIII}{\dot{M}_{\ast,\three}}

\newcommand{\MIII}{M_{\ast,\three}}
\newcommand{\NIII}{N_{\ast,\three}}
\newcommand{\MAR}{\dot{M}_h}
\newcommand{\JLW}{J_{\text{LW}}}

\newcommand{\tauIII}{\tau_{\three}}

\newcommand{\Em}{\mathcal{E}}

\newcommand{\Ec}{\mathcal{E}_c}
\newcommand{\Tc}{\mathcal{T}_c}

\newcommand{\fXII}{f_{X,\two}}
\newcommand{\fXIII}{f_{X,\three}}
\newcommand{\NHMXB}{N_{\text{HMXB}}}
\newcommand{\fsur}{f_{\text{sur}}}

\newcommand{\Mbh}{M_{\bullet}}

\newcommand{\fsc}{f_{\text{sc}}}

\newcommand{\fbh}{f_{\bullet}}

\newcommand{\Mmin}{M_{\min}}
\newcommand{\MminII}{M_{\min,\two}}
\newcommand{\Mmax}{M_{\max}}
\newcommand{\MminIII}{M_{\min,\three}}
\newcommand{\MmaxIII}{M_{\max,\three}}

\newcommand{\rhostardot}{\dt{\rho}_{\ast}}

\newcommand{\rhostardotIII}{\dt{\rho}_{\ast,\three}}

\newcommand{\Nion}{N_{\text{ion}}}
\newcommand{\Nlw}{N_{\text{LW}}}

\newcommand{\fescII}{f_{\text{esc},\textsc{ii}}}
\newcommand{\fescIII}{f_{\text{esc},\textsc{iii}}}
\newcommand{\Msun}{M_{\odot}}
\newcommand{\Tvir}{T_{\text{vir}}}
\newcommand{\Tmin}{T_{\text{min}}}

\newcommand{\dTb}{\delta T_b}

\newcommand{\bin}{\text{bin}}
\newcommand{\fbin}{f_{\bin}}

\newcommand{\sfrunits}{\Msun \ \mathrm{yr}^{-1}}
\newcommand{\intensityunitsnumber}{\text{s}^{-1} \ \text{cm}^{-2} \ \mathrm{Hz}^{-1} \ \text{sr}^{-1}}

\newcommand{\CXRBunits}{\text{erg} \ \text{s}^{-1} \ \text{cm}^{-2} \ \mathrm{Hz}^{-1} \ \text{deg}^{-2}}

\newcommand{\cXunits}{\text{erg} \ \text{s}^{-1} \ (\Msun / \text{yr})^{-1}}
\newcommand{\sfrdunits}{\Msun \ \text{yr}^{-1} \ \text{cMpc}^{-3}}
\newcommand{\LXunits}{\text{erg} \ \text{s}^{-1}}

\newcommand{\slopeunits}{\text{mK} \ \text{MHz}^{-1}}

\title[Global 21-cm Signatures of PopIII]{Unique Signatures of Population III Stars in the Global 21-cm Signal}
\author[Mirocha et al.]{Jordan Mirocha$^1$\textsuperscript{\thanks{mirocha@astro.ucla.edu}}, 
Richard H. Mebane$^1$, 
Steven R. Furlanetto$^1$, 
Krishma Singal$^2$, \newauthor
Donald Trinh$^3$
\\
$^{1}$Department of Physics and Astronomy, University of California, Los Angeles, CA 90024, USA \\
$^{2}$School of Physics, Georgia Institute of Technology, Atlanta, GA
30332\\
$^{3}$Department of Physics and Astronomy, University of California, Irvine, CA 92697, USA \\
}

\begin{document}

\pagerange{\pageref{firstpage}--\pageref{lastpage}} \pubyear{2016}
\maketitle

\begin{abstract}
We investigate the effects of Population III stars on the sky-averaged 21-cm background radiation, which traces the collective emission from all sources of ultraviolet and X-ray photons before reionization is complete. While UV photons from PopIII stars can in principle shift the onset of radiative coupling of the 21-cm transition -- and potentially reionization -- to early times, we find that the remnants of PopIII stars are likely to have a more discernible impact on the 21-cm signal than PopIII stars themselves. The X-rays from such sources preferentially heat the IGM at early times, which elongates the epoch of reheating and results in a more gradual transition from an absorption signal to emission. This gradual heating gives rise to broad, asymmetric wings in the absorption signal, which stand in contrast to the relatively sharp, symmetric signals that arise in models treating PopII sources only. A stronger signature of PopIII, in which the position of the absorption minimum becomes inconsistent with PopII-only models, requires extreme star-forming events that may not be physically plausible, lending further credence to predictions of relatively high frequency absorption troughs, $\nu_{\min} \sim 100$ MHz. As a result, though the trough location alone may not be enough to indicate the presence of PopIII, the asymmetric wings should arise even if only a few PopIII stars form in each halo before the transition to PopII star formation occurs, provided that the PopIII IMF is sufficiently top-heavy and at least some PopIII stars form in binaries.
\end{abstract}

\section{Introduction} \label{sec:intro}
The formation of the first generations of stars in the Universe has been a topic of great interest for several decades \citep[for a recent review, see][]{Bromm2013}. These so-called Population III (PopIII) stars by definition form out of chemically pristine clouds, a fact which is expected to give rise to stellar initial mass functions \citep{Bromm1999, Abel2002b}, atmospheres \citep{Tumlinson2000, Bromm2001, Schaerer2002}, and chemical yields \citep{Heger2002} that are distinct from stars today. Depending on the efficiency with which such stars form in high-$z$ dark matter halos and their longevity as a population, traces of their existence may be found in the reionization history of the intergalactic medium (IGM) \citep[e.g.,][]{Visbal2015,Miranda2017}, chemical abundance patterns in metal-poor stars in the Milky Way \citep[e.g.,][]{Jeon2017,Magg2017}, and pair-instability supernova (PISN) rates at high redshift \citep[e.g.,][]{Whalen2014}. As a result, further investigation of the PopIII epoch continues not only due to a fundamental interest in the physical processes that govern star formation, but because progress in so many other areas may be inextricably linked to the lives and deaths of PopIII stars.

Observationally, PopIII stars have remained elusive. This is not surprising, given that they are expected to form in low mass ($\sim 10^5$-$10^6 \ \Msun$) dark matter halos at the highest redshifts, and probably only in small numbers. There has been only one object detected at high-$z$ with some evidence of PopIII-like stellar population \citep{Sobral2015}, though more ordinary explanations remain viable \citep{Bowler2017}, especially after the recent detection of [CII] \citep{Matthee2017}. Similarly, while there have been a few claims of supernovae at $z\sim 3$ whose properties are consistent with pair-instability models, and thus very massive $M \gtrsim 100 \ \Msun$ supernova progenitors still even after reionization \citep{Cooke2012}, the lack of iron in the most metal-poor nearby stars suggest somewhat less massive progenitors at high-$z$ \citep{Keller2014}.

Another way to constrain PopIII stars is to compare the ionizing photon density of known sources relative to that needed to maintain an ionized IGM at $z \gtrsim 6$, and/or $\tau_e$ values consistent with \textit{Planck}'s latest measurements \citep[e.g.,][]{Visbal2015}. If there is a deficit in the measured photon density relative to what is needed, one might invoke new, as-yet-unseen sources of UV photons at high redshift to close the gap. An analogous but more general argument can be made using the global 21-cm signal \citep{Madau1997,Shaver1999,Furlanetto2006}, since the global 21-cm signal is sensitive to the volume-averaged ionized fraction as well as the thermal history of the IGM and $\Lya$ production histories of galaxies. This was the motivation of \citet{Mirocha2017}, who established a set of global 21-cm predictions calibrated to measurements of the high-$z$ galaxy luminosity function, and included only ``normal'' star-forming galaxies in the model. If these models were to be ruled out observationally, it might indicate the presence of unaccounted for source populations like PopIII stars and their remnants.

The ``deficit elimination'' approach outlined above could provide suggestive evidence of new sources of radiation at high-$z$. However, apparent deficits can vanish if, for example, we simply do not understand the production and/or escape of photons from normal high-$z$ galaxies. Ideally, new sources would provide some unique signature, other than simply adding to the tally of UV and X-ray photons present at $z\gtrsim 6$. 

The zeroth order expectation is of course that the addition of new source populations will boost the luminosity density relative to the ``PopII-only'' predictions of \citet{Mirocha2017}, and thus raise the IGM temperature, ionized fraction, and/or mean $\Lya$ background intensity. However, because the star formation rate density (SFRD) of PopIII stars should be qualitatively different than the SFRD of PopII stars \citep[e.g.,][]{Trenti2009,Xu2016,Mebane2017}, the boost in luminosity density is likely redshift dependent. If strong enough, such a $z$-dependent boost will then manifest as a frequency-dependent modulation of the global 21-cm signal. Our goal in this work is to determine under what circumstances such a modulation can arise, and to determine whether such a signature could be unambiguously associated with PopIII sources (as opposed to, say, uncertainties in the properties of known galaxy populations).

There is certainly no shortage of predictions for the PopIII SFRD in the literature, any one of which we could simply ``plug-in'' to our 21-cm modeling code. However, if we were to simply adopt PopIII star formation histories (SFHs) from the literature, the self-consistency of our model would likely suffer. As a result, we have devised a new toy model that can give rise to a diverse set of histories that builds naturally from our PopII-only models \citep{Mirocha2017}. Such a simple model suffices here, as our goal is \textit{not} to try to bring to bear new insights into the physics governing PopIII star formation, but rather to determine the set of PopIII SFHs that leave the most distinct signature in the global 21-cm signal. 

This work is timely, as several ground-based experiments are currently targeting the global 21-cm signal \citep[e.g., EDGES, BIGHORNS, SCI-HI, SARAS, LEDA;][]{Bowman2010,Sokolowski2015,Voytek2014,Patra2015,Bernardi2016}, with space-based approaches in the design phase \citep[e.g., DARE;][]{Burns2012,Burns2017}. New limits from EDGES and SARAS 2 on the amplitude of the signal \citep{Monsalve2017,Singh2017} are now at the level of many models presented in recent theoretical studies, \citep{Cohen2017,Mirocha2017}, and are thus being tested directly. As a result, the need to invoke new sources may arise sooner rather than later, if such cold reionization scenarios can be ruled out observationally. 

In Section 2 we will describe our model for PopII and PopIII stars, and present their global 21-cm signatures in Section 3. We discuss our results in a broader context in Section 4 before concluding in Section 5.

\section{Models} \label{sec:methods}

\subsection{Population II Star Formation}
Our model for PopII star formation is identical to that used in \citet{Mirocha2017}, so we only summarize here briefly, and defer the interested reader to that paper for more detail. The underlying model is very similar to others appearing in the literature in recent years \citep[e.g.,][]{Mason2015,Sun2016}.

Star formation rates in high-$z$ galaxies are assumed to be directly proportional to the growth rates of their dark matter halos, i.e., $\SFR \propto f_{\ast} \MAR$. We compute the mass growth rates of halos by assuming halos evolve at fixed number density \citep[see, e.g.,][]{Furlanetto2017}, which yields results that are broadly consistent with those found in numerical simulations \citep[e.g.,][]{McBride2009}. We assume the \citet{ShethMoTormen2001} form of the halo mass function, which we compute  using the \textsc{hmf} code \citep{Murray2013}.

We assume that $f_{\ast}$ is a double-power law,
\begin{equation}
    f_{\ast,\two}(M_h,z) = \frac{f_{\ast,0}} {\left(\frac{M_h}{M_{\mathrm{p}}} \right)^{\gamma_{\mathrm{lo}}} + \left(\frac{M_h}{M_{\mathrm{p}}} \right)^{\gamma_{\mathrm{hi}}}} \label{eq:sfe_dpl}
\end{equation}
and calibrate its parameters by fitting to the $z \sim 6$ luminosity function measurements from \citet{Bouwens2015}. Use of, for example, the measurements of \citet{Finkelstein2015} instead results in a $\sim 10$\% difference in the overall normalization of the SFE, which is not enough to qualitatively change our conclusions. Our default case assumes that this SFE is constant in time.

Finally, we note that our default PopII model adopts the \textsc{bpass} version 1.0 single-star models. This results in Lyman-Werner (LW) and Lyman-continuum (LyC) yields of order $\sim 2 \times 10^4$ photons per stellar baryon, much smaller than the yields expected of massive PopIII stars, of order $\Nlw \sim \Nion \sim 10^5$ photons per baryon \citep{Schaerer2002}.

\subsection{X-rays from PopII halos} \label{sec:popII_xrays}
In addition to the double power-law SFE model, the other critical assumption made in \citet{Mirocha2017} was to adopt the empirical relation between X-ray luminosity and SFR ($L_X$-SFR) as found in \citet{Mineo2012a}, with a metallicity-dependence motivated by recent theoretical \citep{Fragos2013} and observational \citep{Brorby2016} findings. 

In the next section we will assume that PopIII X-ray emission analogously tracks the PopIII SFR, but because there is no empirically-calibrated relation, it will be useful to have a simple model to guide us. Readers already familiar with the physical arguments used to explain the $L_X$-SFR relation may skip ahead to \S\ref{sec:popIII_sf}.

As in, e.g., \citet{Mirabel2011}, we estimate the X-ray emission from HMXBs simply as the product of the number of systems in each galaxy, $\NHMXB$, and the typical luminosity of each system, $\mathcal{L}_{\bullet}$, i.e.,
\begin{equation}
    L_X = \NHMXB \mathcal{L}_{\bullet} 
\end{equation}
We assume Eddington-limited accretion with efficiency $\epsilon$, which sets the typical luminosity,
\begin{equation}
    \mathcal{L}_{\bullet} = 1.26 \times 10^{38} \LXunits \left(\frac{\Mbh}{10 \ \Msun} \right) \left(\frac{\epsilon}{0.1} \right) \left(\frac{f_{0.5-8}}{0.84} \right) \label{eq:LHMXB}
\end{equation}
where the factor $f_{0.5-8}=0.84$ above is the fraction of energy emitted in the 0.5-8 keV band for a $10 \ \Msun$ BH with a multi-colour disk (MCD) spectrum \citep{Mitsuda1984}. For an MCD spectrum, we find that the dependence of  $f_{0.5-8 \text{keV}}=0.84$ on BH mass is well approximated (over $10 \leq \Mbh / \Msun \leq 10^6$) by $f_{0.5-8 \text{keV}} \sim \exp\left[-(\Mbh / 1574)^{0.34} \right]$, so even an order of magnitude change in the characteristic mass of remnants has only a mild impact on $L_X$\footnote{Note that our expression is slightly different from that in \citet{Mirabel2011}, as we consider MCD spectra in the 0.5-8 keV band rather than power-law spectra in the 2-10 keV band.}. 

In some cases spectral hardening can be an important consideration, even if the 0.5-8 keV luminosity remains fixed. For example, hardening via attenuation by neutral gas can affect the global 21-cm signal at the $\sim 50$ mK level (for column densities of $\NHI \sim 10^{22} \ \mathrm{cm}^{-2}$), whereas hardening due to the up-scattering of disk photons by a hot corona \citep[e.g., as in the SIMPL model;][]{Steiner2009} is only a minor $\sim 10-20$ mK effect \citep{Mirocha2014}. We neglect such complications\footnote{It could also be that the MCD and SIMPL models are simply not representative of real sources at photon energies $h\nu \lesssim 1$ keV, for which observational constraints are poor.} because, though they may bias inferred $L_X$/SFR values, they are unlikely to change the shape of the 21-cm signal, since the X-ray emission still traces the (continually rising) star formation.

Moving on, we can estimate the number of HMXB systems in a galaxy by assuming a constant SFR, and further assuming that some fraction $\fbh$ of the mass ends up in neutron stars or black holes, of which a fraction $\fbin$ form in binaries, and only a fraction $\fsur$ binaries survive the first supernova. Furthermore, we assume that each system is active for some fraction $f_{\text{act}}$ of its lifetime, $\tau$, which yields
\begin{align}
    N_{\text{HMXB}} & = 20 \left(\frac{\SFR}{\sfrunits} \right) \left(\frac{\fbh}{10^{-3}} \right) \left(\frac{\Mbh}{10 \ \Msun} \right)^{-1} \nonumber \\
    & \times \left(\frac{\fbin}{0.5} \right) \left(\frac{\tau}{20 \ \text{Myr}} \right)  \left(\frac{\fsur}{0.2} \right) \left(\frac{f_{\text{act}}}{0.1} \right) \label{eq:NHMXB}
\end{align}
Here, $\fbh = 10^{-3}$ is the fraction of mass which forms stars with $M_{\ast} > 8 \ \Msun$ assuming a Chabrier IMF. The corresponding figures for Salpeter and Kroupa IMFs are $\fbh = 2 \times 10^{-3}$ and $6 \times 10^{-3}$, respectively. Many of the other factors seem reasonable, but strong arguments beyond the factor of $\sim$ few level are lacking. The combination we have chosen is largely with hindsight, knowing that the canonical normalization of the 0.5-8 keV $L_X$-SFR relation is $c_X \sim 2.6 \times 10^{39} \cXunits$, which the product of Equations \ref{eq:LHMXB} and \ref{eq:NHMXB} roughly yield.

This is a very simple argument, resulting in an equation with many highly uncertain factors. In reality, for example, there is a distribution of luminosities among HMXBs. Though the slope of this distribution function can be explained by fairly simple arguments \citep{Mineo2012a,Postnov2003}, there are still many complex aspects of the HMXB population that we have completely neglected. For example, we have effectively assumed that all systems (when ``on'') reside in the high-soft state (HSS) given our choice of MCD spectrum with no high-energy tail, though of course many known systems are  in the low-hard state. The relative amount of time spent in each -- and why systems transition from one state to the other -- are active areas of research \citep[for a recent review, see, e.g.,][]{Belloni2010}, and we have made no attempt to model these factors. This choice, of pure MCD over a SIMPL spectrum, has little effect on the thermal history \citep{Mirocha2014}, but may be an important distinction if concerned with the $z=0$ unresolved X-ray background (see Section \ref{sec:side_effects}). 

Nonetheless, because the preceding arguments seem to work reasonably well, we employ them again for PopIII star formation in \S\ref{sec:popIII_xrays}. The key point is that if the PopIII IMF is very top-heavy, $\fbh \gg 10^{-3}$,  the PopIII $L_X$/SFR ratio may be larger than that of PopII sources by up to a factor of $\sim 10^3$.

\subsection{Population III Star Formation} \label{sec:popIII_sf}
Even upon changes to the stellar metallicity, $Z$, normalization and $Z$-dependence of the $L_X$-SFR relation, neutral gas contents of galaxies, and evolution of the SFE, \citet{Mirocha2017} found that a deep ($\dTb \lesssim -150$ mK) and late ($\nu \gtrsim 100$ MHz) absorption feature in the global 21-cm signal persisted. This is due to the decline in the SFE in low-mass halos (implied by the LF faint-end slope), which causes the SFRD to also be a steep function of redshift, confining the bulk of UV and X-ray photon production to relatively late times ($z \lesssim 12$). The only exception is if there is a floor in the SFE (steepening in LF faint-end slope), especially if $\Tmin \ll 10^4$ K, which can drive the absorption trough to lower frequencies. 

However, in \citet{Mirocha2017}, we assumed that the $\Tvir < 10^4$ K halos either shared the same properties as the atomic-cooling halos, or did not host star formation at all, both of which are unlikely to be true. Our new models treat these low-mass potentially-PopIII-hosting halos in a simple but physically motivated way that allows their properties to differ from their more massive descendants.  

A minimally-descriptive model of PopIII star formation must explicitly model (or make assumptions about): 
\begin{enumerate}
    \item The minimum mass of PopIII star-forming halos, $\Mmin$.
    \item The star formation history in individual halos during their PopIII phase.
    \item The PopIII stellar initial mass function and atmospheric properties.
    \item The condition which, if met, results in a transition from PopIII to PopII star formation.
\end{enumerate}
All items in this list are coupled, since the minimum mass of PopIII star-forming halos is set by the strength of the LW background \citep{Haiman1997}, which depends on the population-integrated LW emissivity of PopIII halos, which depends on $\Mmin$ and the SFHs of PopIII halos, and so on.

In the following sub-sections we describe these four components in more detail.

\subsubsection{The Minimum Mass}
We begin all calculations at $z_i = 60$, and take $\MminIII$ to be the mass corresponding to halos with virial temperatures of 500 K \citep{Tegmark1997}. The supersonic velocity offset between dark matter and baryons after recombination \citep{Tseliakhovich2010} can modify this initial mass, and thus in principle delay the onset of first-star formation and affect the 21-cm background \citep{McQuinn2012,Fialkov2012}. We will neglect this effect in the present work, as it seems to be relatively minor, at least for the global 21-cm signal \citep{Fialkov2014b}.

\begin{figure}
\begin{center}
\includegraphics[width=0.49\textwidth]{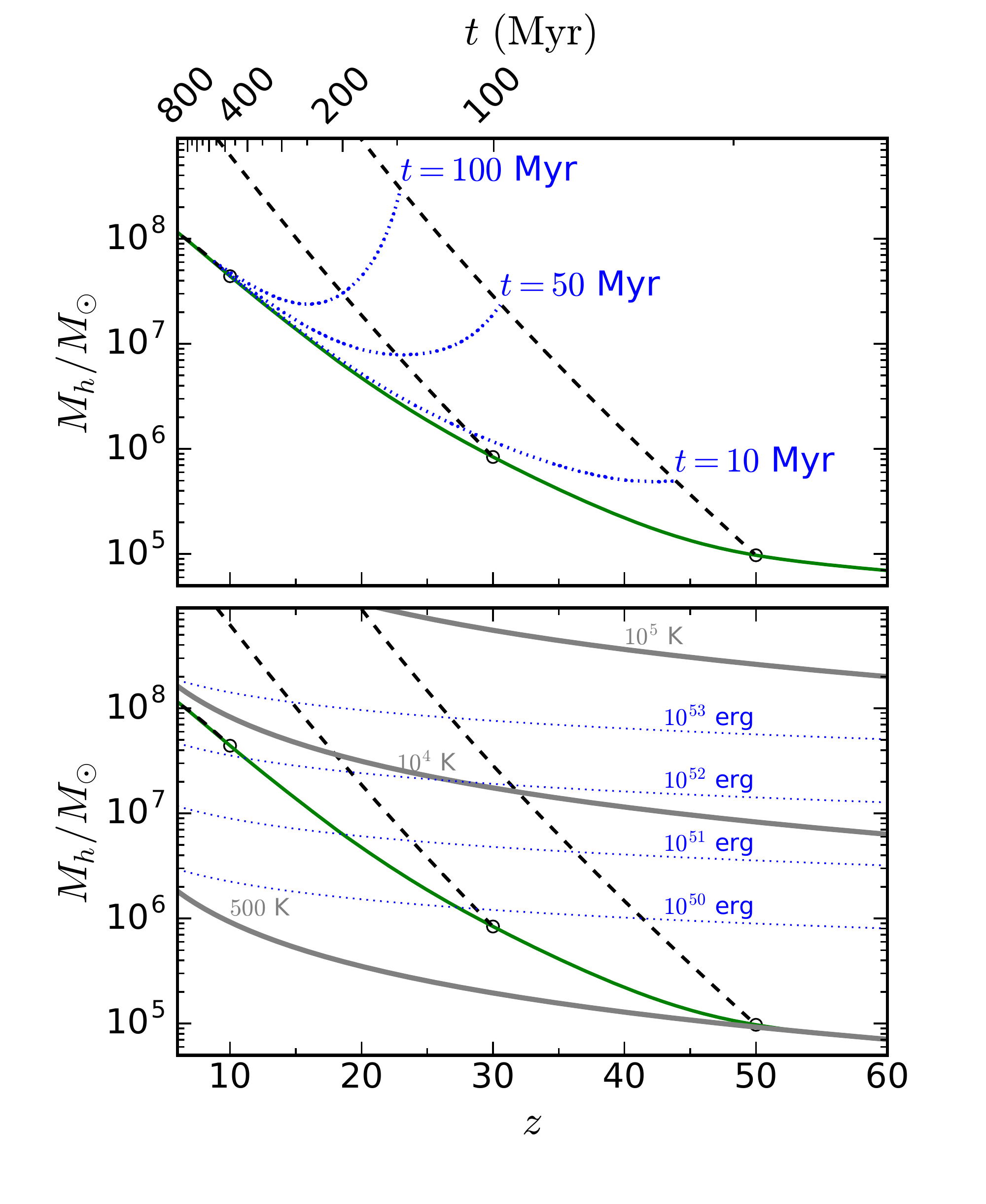}
\caption{Relationships between halo mass, age, and binding energy as a function of redshift. \textit{Bottom:} Lines of constant virial temperature (gray), binding energy (dotted blue), an example minimum mass curve (green), along with the growth histories of halos forming at $z=10$, 20, and 30 (dashed black). \textit{Top:} Lines of constant halo age (dashed blue), as well as the same halo growth histories that are shown in the bottom panel. Note that in both panels, dotted blue lines highlight the means by which we trigger the transition from PopIII to PopII star formation.}
\label{fig:trajectories}
\end{center}
\end{figure}

Once PopIII stars begin forming, they generate a diffuse LW background (with mean intensity $\Jlw$) that can in principle globally regulate star formation by raising the minimum mass threshold \citep{Haiman1997}. To include this effect, we assume the $\Mmin(\Jlw)$ relation of \citet{Visbal2014} (their Equation 4), and solve iteratively for $\MminIII$ and $\Jlw$, requiring that the PopIII SFRD has converged to a relative accuracy of better than 5\% at all redshifts before computing the global 21-cm signal. Typically, only $\sim 5$-10 iterations are required in order to reach convergence\footnote{Scenarios with very efficient PopIII star formation can drive rapid evolution in $\Mmin$, which can lead to the need for many iterations ($\sim$ dozens) to reach convergence. To reduce computational time, we have found that simply averaging subsequent solutions for $\Mmin$ every $\sim 5$ iterations damps out sharp features that arise in $\Mmin$ solutions under these circumstances.}. An example $\Mmin(z)$ curve is shown in green in Figure \ref{fig:trajectories}, along with relationships between halo age and binding energy, to be discussed further in the \S\ref{sec:transition}.

But first, because the detailed properties of PopIII stars govern the strength of the diffuse LW background and thus $\Mmin$, we focus on them next in \S\ref{sec:SFHIII}.

\subsubsection{PopIII Star Formation Histories} \label{sec:SFHIII}
We assume that each PopIII star-forming halo has a single star-forming region \citep[e.g., ][]{OShea2007}. We assume also that each episode of star formation produces the same mass in PopIII stars on average, which will write as the product of the typical number of stars and their characteristic mass, $\NIII \MIII$, to build intuition in what follows. Given a typical lifetime of $\tau_{\three}$ and a ``recovery time'' between star-forming episodes\footnote{Note that numerical simulations suggest that the recovery time could be as short as 10-20 Myr \citep[e.g.,][]{OShea2005}, or as long as a Hubble time, depending on PopIII stellar mass \citep{Jeon2014}, which would drive $\SFRIII$ to values lower than $\sim 10^{-5} \ \sfrunits$. However PopIII stars may also form in larger numbers with a spectrum of masses. So the $\SFRIII \sim 2 \times 10^{-5} \sfrunits$ quoted here should only be considered a rough estimate.}, $\tau_{\mathrm{recov}}$, the mean SFR in PopIII halos can be written as
\begin{equation}
    \SFRIII = \frac{\NIII \MIII}{\tauIII + \tau_{\mathrm{recov}}} \label{eq:SFRIII}
\end{equation}
If PopIII stars are massive ($\MIII \sim \ 100 \Msun$) and form in isolation ($\NIII = 1$), with a typical lifetime of order $\tauIII \sim 5$ Myr and no recovery time (i.e., they form one after the next), then $\SFRIII = 2 \times 10^{-5} \ \sfrunits$. Compared to PopII halos, in which $\SFRII \propto M_h^{5/3} (1+z)^{5/2}$ (roughly), PopIII halos form stars much less efficiently. For example, if we define a PopIII SFE that is analogous to the PopII SFE (i.e., relative to halo growth rate) for a rough comparison, then $f_{\ast,\textsc{iii}} \sim M_h^{-1} (1+z)^{-5/2}$. This scaling ignores the potential connection between, for example, the recovery time and mass accretion rates, but serves to illustrate the qualitative difference between the efficiency of PopII and PopIII star formation in our model. In practice, we do not choose each quantity in Equation \ref{eq:SFRIII} separately, but instead  treat $\SFRIII$ itself as our main free parameter\footnote{For individual halos this SFR is ill-defined, since the `true' history is a series of discrete bursts. It should be treated as a population-averaged quantity.}. 

\begin{figure*}
\begin{center}
\includegraphics[width=0.98\textwidth]{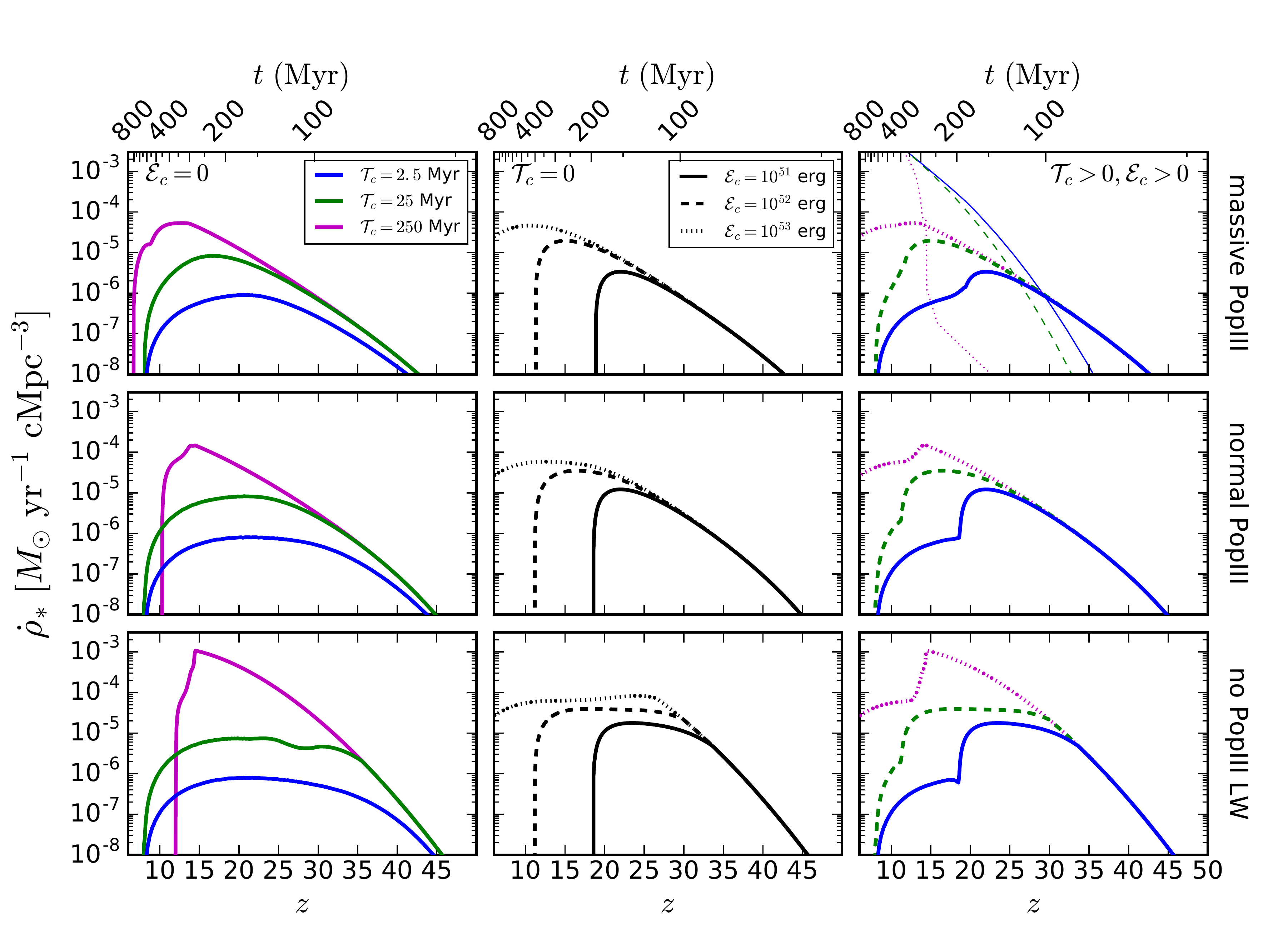}
\caption{Models for the PopIII star formation rate density, assuming a constant $\SFRIII = 10^{-5} \ \sfrunits$ in all PopIII halos. From left to right, we explore the different transition mechanisms, starting with time-limited PopIII star formation (left), binding-energy-limited (middle), and a model in which the both mechanisms are at work (right). Each row adopts a different model for the PopIII stellar properties, from the massive ($120 \ \Msun$) case, which sets $\Nlw \sim 10^5$ (top), to no LW emission at all (bottom). Example PopII star formation histories are shown in thin lines in the upper right panel. Note that the inferred SFRD at $z \sim 6$ is of order $\sim$ few $\times 10^{-2} \ \sfrdunits$.}
\label{fig:popIII_sfrd}
\end{center}
\end{figure*}

The assumption of a single site of star formation results in a PopIII SFRD that is proportional to the number of PopIII halos between the minimum mass, $\Mmin$, and some maximum mass, $\Mmax$, that is determined by our transition criteria, i.e.,
\begin{equation}
    \rhostardotIII =  \SFRIII \int_{\MminIII}^{\MmaxIII} \frac{dn}{dm} dm .
\end{equation}
Here, $dn/dm$ is the mass function of dark matter halos. We assume that PopII star formation commences immediately after the PopIII epoch, i.e., $\MminII \equiv \MmaxIII$. We set $\MmaxIII$ via simple arguments outlined in the next subsection, and shown graphically in Figure \ref{fig:trajectories}.

In most of our models, the total number density of PopIII star-forming halos is $\sim 1-10 \ \mathrm{cMpc}^{-3}$ at the peak of PopIII star formation, which is why the PopIII SFRD generally peaks at values of order $\sim \SFRIII \ \mathrm{cMpc}^{-3}$ (as shown in Figure \ref{fig:popIII_sfrd}), which we will see in detail in the next section.

\subsubsection{Properties of PopIII Stars}
For the bulk of this study, we assume that PopIII stars are massive, $\sim 100 \ \Msun$ stars. We derive UV photon yields for PopIII stars from \citet{Schaerer2002} (the time-averaged properties in his Table 4). Our reference model assumes $120 \ \Msun$ stars, which emit $1.4 \times 10^{50}$ Hydrogen-ionizing photons per second, and $1.6 \times 10^{50}$ LW photons per second averaged over their $2.5$ Myr lifetime. For stars more massive than $60 \ \Msun$, lifetime-integrated yields change by only $\sim 10$\%, since the lower mass stars are less luminous but also live longer. It is these lifetime-integrated values for LyC and LW photon production, commonly denoted with $\Nion$ and $\Nlw$, respectively, that we use to convert star formation rates to LyC and LW photon production rates throughout.

\subsubsection{Transition to PopII Star Formation} \label{sec:transition}
We assume that PopIII stars continue to form at rate $\SFRIII$ until some set of criteria are met that trigger the transition from PopIII to PopII star formation. In numerical simulations, a critical metallicity \citep[$Z_c \sim 10^{-3.5}$][]{Bromm2003}, determines whether a cell forms PopII or PopIII stars. However, because our calculations have no spatial information, we cannot impose the transition based on local gas conditions (we only have global halo properties at our disposal). \citet{Mebane2017} find that use of the \textit{mean} halo metallicity, in lieu of any local metallicity information, can result in PopIII SFHs that resemble those of more sophisticated calculations. We adopt a simpler approach in this work.

We introduce two parameters that aim to bound cases in which the transition to PopIII is triggered either (i) once halos have generated a sufficient metal mass, and/or (ii) are sufficiently massive that metals produced in supernovae remain gravitationally bound to the halo. The first case is obtained by assuming halos form PopIII stars for a given amount of time, $\Tc$, which in our framework is equivalent to assuming the PopIII phase always results in the same mass in stars or metals (since we assume a constant PopIII SFR in individual halos). The second limit is obtained by assuming PopIII halos transition to PopII star formation once their binding energy exceeds some critical value, $\Ec$.

Large values of $\Ec$ represent cases in which the supernovae of the first stars are so energetic that material in halos with $\Em < \Ec$ would become completely unbound. Smaller values of $\Ec$ represent cases of less energetic supernovae, which even relatively small halos can sustain. In reality, there is likely a link between $\Tc$ and $\Ec$, since the supernovae that produce the most metals are also probably the most energetic. In other words, halos with and without PISN may still have comparable PopIII SFHs, since efficient metal production in PISN may be counteracted by efficient metal expulsion. We make no attempt to model this in detail, though such cases do lie within our model grid (see \S\ref{sec:popIII_gs_grids}).

Model halo mass growth histories are shown in Figure \ref{fig:trajectories} for halos that cross the minimum mass threshold at different times, with lines of constant halo virial temperature, binding energy, and age indicated for reference. The minimum mass curve in this figure is set assuming $\Tc=2.5$ Myr, $\Ec = 10^{51}$ erg, and $\SFRIII=10^{-5} \ \sfrunits$. Note that the trajectory of a halo forming at $z\sim 10$ in this model is roughly parallel to the minimum mass curve (due to the decline in halo accretion rates), meaning there will be no new PopIII halos at $z \lesssim 10$. 

When operating in isolation, each transition mechanism produces qualitatively similar PopIII SFRDs, as shown in the left and center columns of Figure \ref{fig:popIII_sfrd}. Both lead to rising SFRDs at early times, which eventually peak (or plateau) before tending to zero at late times (in most cases). The most noticeable difference between the $\Tc$-limited (left) and $\Ec$-limited (middle) models is that the latter can produce very sharp PopIII SFRDs which rise and fall rapidly at early times. Note that similar histories can also occur if $\Mmin$ rises above the atomic-cooling threshold at early times \citep[if, e.g., PopII star formation is very efficient and generates a strong LW background;][]{Mebane2017}, enabling the formation of metal-free stellar populations with a normal IMF. Such stars are still technically PopIII, though for our purposes, and in \citet{Mebane2017}, it is appropriate to count such halos in the PopII SFRD since it is the most massive stars (and their remnants) to which the global 21-cm signal and PISN rates are most sensitive. We will revisit this point in \S\ref{sec:popIII_gs_basics}, as such histories have a particularly distinct impact on the global 21-cm signal.

If $\Tc$ and $\Ec$ are both non-zero, they can combine to produce multi-component SFRDs, the features of which are often sharp given the simplicity of our model (right column of Figure \ref{fig:popIII_sfrd}). Though likely unrealistic, we make no effort to smooth out such features as they make it easy to visually identify when the PopIII SFRD is governed by $\Tc$ or $\Ec$, both in the SFRD itself, as well as the global 21-cm signal (see \S\ref{sec:results}). If the values are just right, the SFRDs can be smooth, and sometimes very nearly flat.

These results are subject to the assumed production efficiency of LW photons, $\Nlw$, which our default ``massive PopIII'' case assumes is $\Nlw \sim 10^5$, as is expected to be the case for $\sim 100 \ \Msun$ metal-free stars \citet{Schaerer2002}. The middle and bottom rows of Figure \ref{fig:popIII_sfrd} change $\Nlw$ and thus the strength of LW feedback. If LW emission from PopIII stars is reduced, by assuming their properties are comparable to metal-poor stars with a Salpeter IMF (``normal IMF''), PopIII stars can form in lower mass halos, and boost the overall SFRD. In the extreme limit of no PopIII LW emission, PopIII star formation is regulated by the PopII-generated LW background, and can result in strong, very flat PopIII SFHs. 

This differences between $\Tc$ and $\Ec$ models can be better understood by referring back to Figure \ref{fig:trajectories}. For $\Tc$-limited models, the $\Mmin$ and $\Mmax$ curves are quasi-parallel, since the mass range depends on the growth history of halos (i.e., $\MminIII$ at $z$ sets $\MmaxIII$ at $z^{\prime} < z$). The same is not true of the $\Ec$-limited models, whose $\MmaxIII$ contours can be drawn without reference to the growth of individual halos. As a result, the interface between $\MminIII$ and $\MmaxIII$ can be sharp (the binding energy is only linear in redshift, while $\Mmin$ is typically steeper), and PopIII SFRDs can rapidly decline to zero as $\MminIII$ overtakes $\MmaxIII$. In other words, there comes a point for $\Ec$-limited models in which \textit{all} halos suddenly satisfy the transition criterion, $\Em \geq \Ec$, which means they immediately commence PopII star formation, even if they have yet to form PopIII stars. Note that in this case, if the critical $\Ec$ curve is comparable to the atomic cooling threshold, the stars formed could follow a normal IMF despite still being metal free.

In the $\Ec=0$ limit, such sharp SFRDs never occur. PopIII star formation will continue globally at some non-zero level until the minimum mass is growing more rapidly than a halo of the minimum mass. In other words, some halos will be growing too slowly to ever `catch up' with $\MminIII$ and will never be able to form stars at all. In our models, this typically occurs at $z \lesssim 10$ (see Figure \ref{fig:trajectories}, in which a halo forming at $z=10$ grows at the same rate as $\MminIII$). As in \citet{Mebane2017}, this is what ultimately drives PopIII stars to extinction in most models.

\subsection{X-rays from PopIII halos} \label{sec:popIII_xrays}
We also assume that halos hosting PopIII stars emit X-rays. In general, the emission could come from inverse Compton scattering in supernova remnants, or from bremmstrahlung as ISM gas cools. As our reference case, we assume that some fraction of PopIII stars will form in binaries, and ultimately produce an X-ray binary system, which renders our PopIII model completely analogous to the PopII model (see \S\ref{sec:popII_xrays}), except for its susceptibility to global LW feedback, and finite lifetime imposed by $\Tc$ and $\Ec$. 

As a result, for PopIII halos we introduce a scaling factor for the PopIII $L_X$-SFR relation, $f_{X,\three}$, defined relative to the canonical PopII $L_X$-SFR relation, i.e.,
\begin{equation}
    L_{X,\three} = f_{X,\three} \times 2.6 \times 10^{39} \ \cXunits
\end{equation}
Whereas the local (PopII) $L_X$-SFR relation, probably does not grow by more than a factor of $\sim 10$ as metallicity decreases, Equations \ref{eq:LHMXB} and \ref{eq:NHMXB} tell us that PopIII sources may be $\sim 10^3$ times more efficient at producing X-rays per unit stellar mass formed, if indeed their IMF is top-heavy (i.e., $\fbh \gg 10^{-3}$).

Though the typical remnant mass may be larger for PopIII HMXBs, and thus reduce the value of $f_{0.5-8}$ and perhaps also $\tau$, these are likely fairly modest, factor of $\sim$ few, effects. As a result, it is not difficult to imagine a scenario in which PopIII sources are substantially more efficient at producing X-rays than PopII halos, so long is $\fbh$ is large. If indeed  $\fXIII > 1$, the X-ray background will remain dominated by PopIII even after the total SFRD becomes dominated by PopII sources, the latter of which typically occurs at $15 \lesssim z \lesssim 30$ (see upper right corner of Figure \ref{fig:popIII_sfrd}).

Note that we will not explicitly model the binary fraction or PopIII IMF, since $\fXIII$ is also degenerate with $\SFRIII$. We will thus attempt to qualify expectations for constraining $\fXIII$ and $\SFRIII$ accordingly throughout.

For the rest of this paper we assume that the minimum mass is set only by $\Jlw$, though in principle $\Mmin$ could depend on the X-ray background as well, since X-rays can boost the electron fraction and thus catalyze $H_2$ formation in dense clouds \citep[e.g.,][]{Machacek2003,Glover2016,Ricotti2016}. We will revisit this potential complication in Section \ref{sec:discussion}.

\subsection{Generating the Global 21-cm Signal}
All calculations were conducted with the \ares\ code\footnote{https://bitbucket.org/mirochaj/ares; v0.4}. \ares\ treats the IGM as a two-phase medium, tracking separately the growth in the volume filling factor of ionized gas and the mean temperature and ionization state of gas in the ``bulk'' IGM beyond. That is, \ares\ does \textit{not} generate a three-dimensional realization of the 21-cm field using, e.g., semi-numeric techniques \citep{Mesinger2011}. For a more detailed description of how \ares\ solves for the ionization and thermal histories, see Section 2 of \citet{Mirocha2014}. 

\ares\ outsources a few important calculations to well-established software packages. For example, we generate initial conditions for the state of the high-$z$ IGM using the \textsc{CosmoRec} code \citep{Chluba2011}, and halo mass functions using the \textsc{hmf-calc} code \citep{Murray2013}, which itself depends on the \textit{Code for Anisotropies in the Microwave Background} \citep[\textsc{camb};][]{Lewis2000}. 

A few additional notes are warranted regarding various atomic physics calculations that occur within \ares.

We compute the rate of collisional excitation/de-excitation of the hyperfine states using the tabulated values in \citet{Zygelman2005} and take the radiative coupling coefficient \citep{Wouthuysen1952,Field1958} to be $x_{\alpha} = 1.81 \times 10^{11} \widehat{J}_{\alpha} S_{\alpha} / (1 + z)$, where $S_{\alpha}$ is a factor of order unity that accounts for line profile effects \citep{Chen2004,FurlanettoPritchard2006,Chuzhoy2006,Hirata2006}, and $\widehat{J}_{\alpha}$ is the intensity of the $\Lya$ background in units of $\intensityunitsnumber$. We adopt the formulae from \citet{FurlanettoPritchard2006} to compute $S_{\alpha}$.

In addition, we follow \citet{Pritchard2006} in computing the fraction of $\Lyn$ photons that cascade through the $\Lya$ resonance, and use the lookup tables of \citet{Furlanetto2010} to determine the fraction of energy that photo-electrons deposit as heat, further ionization, and excitation in the gas. We use the fits of \citet{Verner1996} for bound-free absorption cross-sections, and adopt the formulae for recombination and cooling rate coefficients from the appendices of \citet{Fukugita1994}.

We adopt \textit{Planck} cosmological parameters \citep{Planck2015} throughout.

\section{Results} \label{sec:results}
In this section, we focus on how PopIII stars affect the global 21-cm signal. First, we explore a small set of realizations to build some intuition (\S\ref{sec:popIII_gs_basics}). Then, we move on to an expanded grid of models, and attempt to determine if the PopIII-induced modulations of the signal are generically distinct from those brought about by variations in the parameters governing PopII star formation (\S\ref{sec:popIII_gs_grids}). To close the section, we examine whether any of our PopIII scenarios are in tension with pre-existing measurements (\S\ref{sec:side_effects}).

\subsection{PopIII Signatures: Basic Features} \label{sec:popIII_gs_basics}
\begin{figure*}
\begin{center}
\includegraphics[width=0.98\textwidth]{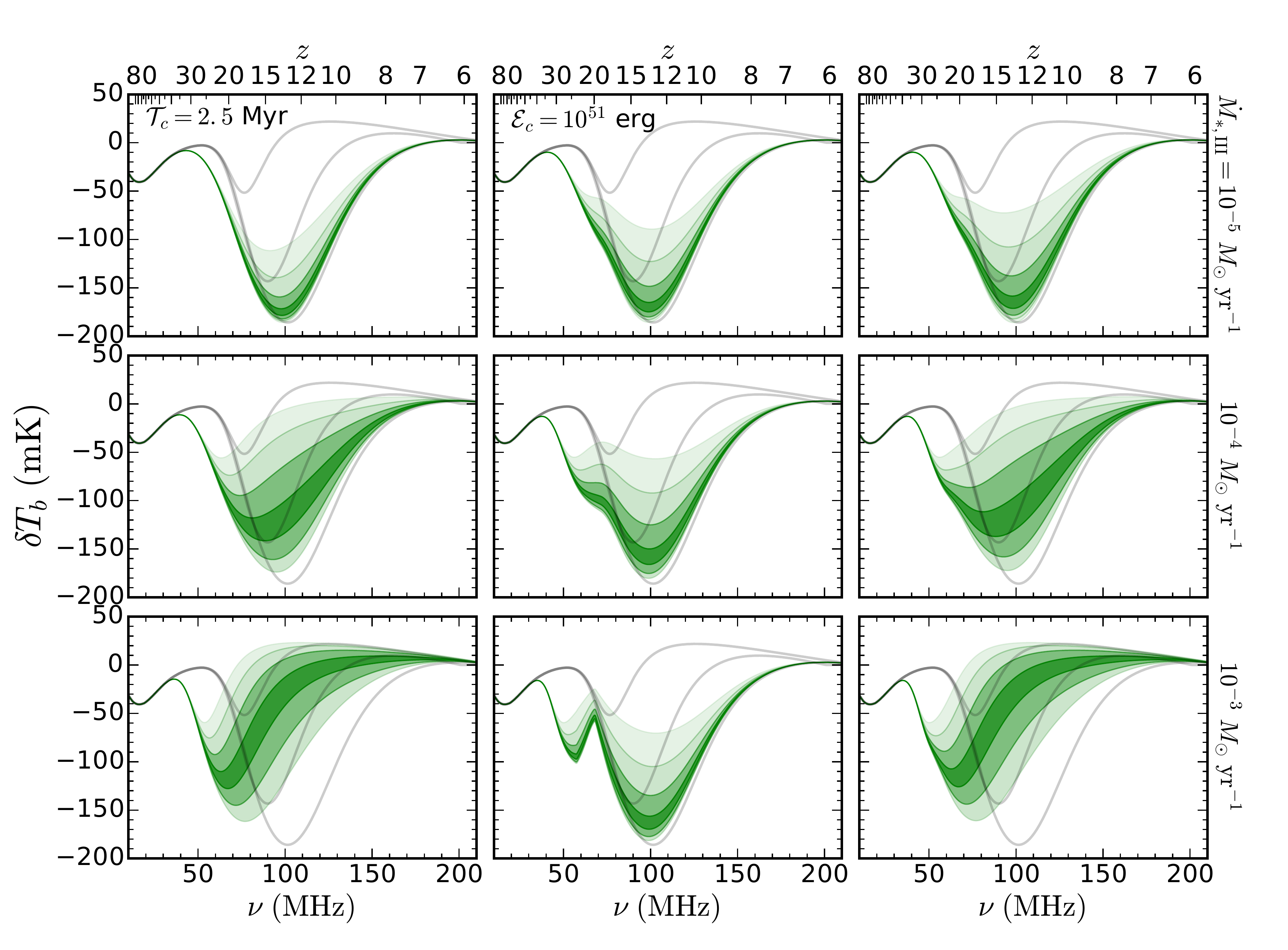}
\caption{Global 21-cm signal realizations with PopIII sources. From top to bottom, each row assumes increasingly efficient PopIII star formation in individual halos, while each column explores a different transition mechanism, including pure time-limited cases ($\Tc=2.5$ Myr; left), pure binding-energy-limited cases ($\Ec=10^{51}$ erg; middle), and a case with both effects in operation (right; colours and line-styles correspond to first two columns). The width of each coloured band corresponds to a factor of 32 change in $\fXIII$ over the interval ($25 \leq \fXIII \leq 3200$), with individual factor of 2 changes indicated by opacity. For example, the most opaque band corresponds to $200 \leq \fXIII \leq 400$. Three example PopII-only realizations are shown in gray in each panel, each with increasingly efficient heating, which drive both weaker troughs and stronger emission features.}
\label{fig:popIII_gs_XRIII}
\end{center}
\end{figure*}

In Figure \ref{fig:popIII_gs_XRIII}, we show several realizations of the global 21-cm signal assuming different models for PopIII stars, but holding the PopII component of the model fixed. In each row, we assume a different PopIII SFR per halo, $\SFRIII$, while each column adopts a different transition mechanism, including time (left), binding energy (middle), and a scenario in which both of these mechanisms are at work, and correlated with each other (right). The width of each semi-transparent band corresponds to a factor of 2 change in $f_{X,\three}$, from $25 \leq \fXIII \leq 3200$. The lower limit of 25 was chosen because it is usually the point when the trough depth first comes to be visibly affected by PopIII sources.  The upper limit of $3200$ does not correspond directly to a special case, though is about the limit of what one would expect for a source population of exclusively $100 \ \Msun$ stars that all form in binaries (see Eqs. \ref{eq:LHMXB}-\ref{eq:NHMXB}). For reference, the most opaque band corresponds to $200 \leq f_{X,\three} \leq 400$.

Each panel also shows three example PopII-only models with increasingly efficient X-ray heating. The model with the deepest trough is our reference PopII-only model \citep[referred to as the \dpl\ model in ][]{Mirocha2017}, while the intermediate case assumes low metallicity star-forming regions and a strong connection between $L_X$-SFR and $Z$ (as $L_X \propto Z^{-0.6}$), which boost $L_X$/SFR by a factor of 9. Unless there are much more efficient X-ray sources at high-$z$, these realizations roughly span the range of expected trough depths. The model with the weakest trough adds a new population of X-ray sources with a soft (unabsorbed) $\alpha_X=-1.5$ power-law spectrum that produces 0.5-8 keV photons 10 times more efficiently than our default model, i.e., with $L_X/\mathrm{SFR}=2.6 \times 10^{40} \ \cXunits$. 

The top row of Figure \ref{fig:popIII_gs_XRIII} is most similar to the `classic' conception of PopIII as isolated, massive, short-lived stars, since $\SFRIII \sim 10^{-5}$ corresponds to a single $100 \ \Msun$ star forming every 10 Myr (via Eq. \ref{eq:SFRIII}). In this case, the frequency of the absorption minimum is hardly affected, though its amplitude can be reduced if $\fXIII \gg 1$. If the binding energy is an important factor in the transition to PopII star formation (center and right columns), the low-frequency tail of the absorption trough grows slightly, as PopIII star formation is effectively allowed to persist for longer times in very high-$z$ halos. 

At this stage, the \textit{absolute} depth of the absorption trough appears not to be a powerful discriminant between PopII+PopIII models (green) and PopII-only models (gray), as considerable uncertainty (to be explored further in the next section) in PopII sources still remains. However, the depth of the trough \textit{relative} to the emission peak, and the timing between extrema of the signal, appear to have more potential. For example, even as the trough becomes shallower in PopIII models, the emission maximum remains largely unchanged. Such behavior is not seen in PopII-only models, in which reductions in the trough amplitude give rise to stronger emission features. This is because heating of the IGM is driven by sources whose SFRD rises monotonically with time, whereas PopIII sources provide ``extra'' heating only at early times. This decoupling of the amplitudes of the minimum and maximum leads to very broad, asymmetric absorption signals. 

This general trend continues as we increase $\SFRIII$ in the middle and bottom rows of Figure \ref{fig:popIII_gs_XRIII}. In extreme cases, a second absorption trough appears (center column, bottom two rows). Such behavior is probably unlikely, as it requires strong ($\SFRIII \gtrsim 10^{-4} \ \sfrunits$) but globally short-lived ($z \gtrsim 20$ only) PopIII star formation, so that the UV background generated by PopII sources is not strong enough to maintain the Wouthuysen-Field coupling after PopIII stars die out. In our model, this is only achieved if $\Tc \sim 0$, since halos can effectively skip the PopIII phase if their binding energy already exceeds $\Ec$ when they cross the minimum mass threshold. Even a single, short-lived episode of PopIII star formation can prevent the emergence of a double trough\footnote{Another way for halos to skip the PopIII phase if metals expelled by neighboring halos are accreted before stars form \citep{Smith2015}, though it seems unlikely that such a process could affect enough halos at early times to completely eliminate PopIII star formation globally given the generally low volume-filling fraction of metals seen in simulations \citep[e.g.,][]{Jaacks2017}.} (right column, bottom two rows). The binding energy alone is probably not a sufficient criterion for triggering the transition to PopII star formation, as it does not influence the cooling properties of gas in high-$z$ halos\footnote{Except if that critical binding energy is $\sim 10^{51.5} - 10^{52}$ erg at $z \gtrsim 30$, which are comparable to the atomic cooling threshold, though with milder redshift evolution. See the bottom panel of Figure \ref{fig:trajectories}.}. Note that here we have assumed massive PopIII stars, so the SFR values explored from top to bottom should really be interpreted as the product $\SFRIII (\Nlw/10^5)$, which implies that PopIII sources with a more normal IMF (and $\Nlw \sim 10^4$) will have a less dramatic effect on the 21-cm signal. We will revisit this point in \S\ref{sec:dramatic_departures}.

Given the observational implications of a double trough, the likelihood of a rapid rise and fall in the PopIII SFRD warrants further discussion, which we also defer to \S\ref{sec:dramatic_departures}. 

\subsection{Are Broad Asymmetric Troughs a Generic Feature of PopIII?} \label{sec:popII_vs_popIII}
In the previous section, we showed a small set of representative PopIII SFRDs and the corresponding global 21-cm spectra for select values of $\Tc$ and $\Ec$. Our goal in the rest of the paper will be to determine if broad asymmetric troughs are a feature of PopIII models that is (i) distinguishable from PopII-only models, and (ii) expected for a broader range of PopIII models.

\subsubsection{A Large Set of PopII-only Models} \label{sec:popII_mc}
\begin{figure*}
\begin{center}
\includegraphics[width=0.98\textwidth]{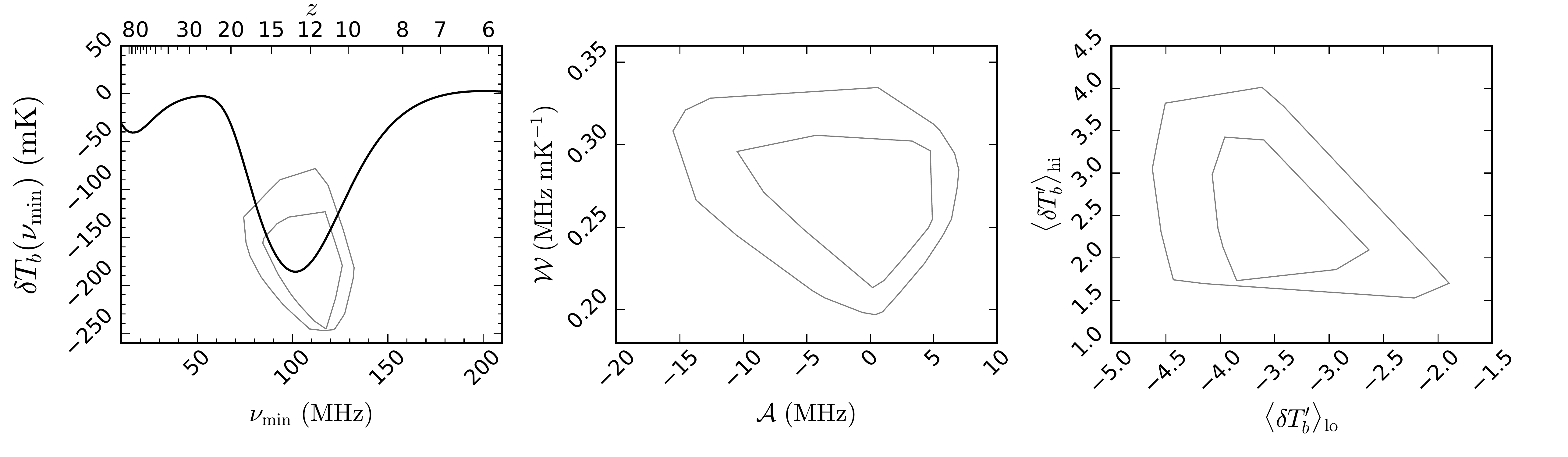}
\caption{Distillation of the \citet{Mirocha2017} models. \textit{Left:} Fiducial PopII-only model (solid black curve), with contours enclosing the location of the absorption trough in all realizations (largest polygon), and a refined subset (smaller polygon). \text{Middle:} Relationship between the asymmetry of the signal, $\hwhmdiff$ (Eq. \ref{eq:asymmetry}),  and the prominence of its wings, $\squash$ (Eq. \ref{eq:squash}). \textit{Right:} Relationship between the mean slope of the global 21-cm signal, as measured at frequencies between the first extremum and the trough, $\meanslopelo$, and at frequencies between the absorption trough and emission peak, $\meanslopehi$.}
\label{fig:popII_soa}
\end{center}
\end{figure*}

To determine if the presence of PopIII can be detected despite uncertainties in our understanding of PopII galaxies, we generate a superset of the \citet{Mirocha2017} models ($N=10^5$ of them) by Monte Carlo sampling the parameter space defined in Table \ref{tab:popII_parameters} and shown graphically in Figure \ref{fig:popII_soa}. 

The bounds of this parameter space are chosen to encompass:
\begin{itemize}
    \item The range of single-star spectral models in \textsc{BPASS} version 1.0 \citep{Eldridge2009} over the entire metallicity range, $0.001 \leq Z \leq 0.04$. Use of the version 1.0 \textsc{BPASS} models is to maintain consistency with the \citet{Mirocha2017} models, but note that $\beta_X$ (see next bullet) has a larger impact than $Z$ itself, given that we force our models to match the galaxy LF \citep[see Figure 5 and associated text in][]{Mirocha2017}
    \item Uncertainties in the $z=0$ $L_X$-SFR relation (factor of 5), and its dependence on metallicity ($L_X \propto Z^{\beta_X}$), from $Z$-independent all the way up to $\beta_X = -0.8$ \citep[slightly steeper than the preferred value in][]{Brorby2016}.
    \item Uncertainty in the redshift evolution of the SFE of PopII halos, which we allow to evolve as a power-law with index $\gamma_x$, between $f_{\ast} \propto (1+z)^{-1}$ and $f_{\ast} \propto (1+z)^{1}$. Simple models suggest a mild $(1+z)^{1/2}$-$(1+z)$ dependence \citep{Furlanetto2017}, but observations at high-$z$ are roughly consistent with all of these scenarios \citep{Sun2016,Mirocha2017}.
    \item Uncertainty in the faint-end slope of the LF, which we model allowing the low-mass SFE to steepen as $f_{\ast}(M_h) = \bigg[1 + \left(2^{\mu / 3} - 1\right) \left(\frac{M_h}{M_c} \right)^{-\mu} \bigg]^{-3 / \mu}$, as in \citet{OShea2015}. We take $\mu=1$ and vary $M_c$ as a free parameter.
    \item Uncertainty in the escape of X-ray photons, as parameterized by a characteristic neutral column density, $\NHI$, in a range consistent with what is seen (at least in low-mass $10^7-10^8 \Msun$ halos) in simulations \citep{Das2017}. We include the opacity of neutral hydrogen and helium, but neglect contributions from HeII and metals.
\end{itemize}

\begin{table}
\begin{tabular}{ | l | l | l }
\hline
name & full range & refined range \\
\hline
$Z$         & $[0.001, 0.04]$ & $[0.001, 0.003]$ \\
\hline
$\gamma_z$ & [-1, 1] & [-0.1, 0.1]  \\
$M_c / M_{\odot}$ & [$10^7$, $10^{11}$] & [$10^7$, $10^{9}$] \\
\hline
$c_X$       & $[1, 5] \times 10^{39} \ \cXunits $ & \\
$\beta_X$   & $[-0.8,0]$ & $[-0.7,-0.5]$ \\
\hline
$\NHI$ & $[10^{19}, 10^{22}] \ \mathrm{cm}^{-2}$ & \\ 
$f_{\mathrm{esc},LyC}$ & $[0.01, 0.3]$ & 
\end{tabular}
\caption{Parameter space surveyed for PopII models. Each row provides information for a different free parameter (introduced in text), with the range of values surveyed first for the most conservative set (middle), and then for a more restrictive set as described in the text (right).}
\label{tab:popII_parameters}
\end{table}

The basic properties of this model set are shown in Figure \ref{fig:popII_soa}. In the left panel, we show the reference model from \citet{Mirocha2017}, in addition to contours bounding the location of the absorption minimum in all models in our MC-generated set (see Table \ref{tab:popII_parameters}). As expected, models are roughly centered on our reference PopII-only, with $\nu_{\min} \sim 100$ MHz and $\dTb(\nu_{\min}) \sim -180$ mK\footnote{Due to a bug in the calculation of $S_{\alpha}$ in a previous version of \textsc{ares}, troughs in these models occur slightly ($\sim$ few MHz) earlier than those in \citet{Mirocha2017}.}. Note that our reference PopII-only model is similar to other models in the literature, for example the \textsc{bright galaxies} model of \citet{Mesinger2016}.

Models with deeper troughs at higher frequencies than those in our reference PopII-only model have harder X-ray spectra, less efficient star formation, and/or less efficient X-ray production per unit SFR (e.g., $\beta_X \sim 0$ or small $c_X$). Shallower and earlier troughs occur when $Z \ll Z_{\odot}$, when the $L_X$-SFR relation depends strongly on $Z$ ($\beta_X \ll 0$), and or when the intrinsic absorption in host galaxies is minimal. We find that strong neutral absorption (at the level of $\NHI \sim 10^{21.5} \ \mathrm{cm}^{-2}$) almost exactly counteracts the more efficient heating caused by low-$Z$ boosts to the $L_X$-SFR relation, and results in global 21-cm spectra consistent with the pure PopII-only model.

Our most conservative bounds on PopII models are represented by the largest black contour, with a more refined set of models bounded by the inner contour, that assumes that the SFE of PopII halos does not evolve with time, that high-$z$ galaxies are low metallicity ($0.001 \leq Z \leq 0.003$), and that low $Z$ is reflected both in the UV spectrum of sources and their X-ray emissions (by limiting $-0.7 \leq \beta_X \leq -0.5$, i.e., assuming the \citet{Brorby2016} best-fit is correct). There are simple theoretical arguments in support of each refinement, though current measurements do not require such revisions to the model. For that reason, the inner contours are meant to indicate how progress in the coming years, e.g., better constraints on the redshift evolution of the LF and metallicities of very high-$z$ galaxies, might feed back into the calibration of our PopII-only models and thus enhance our sensitivity to PopIII sources.

In the center panel, we focus on two new metrics of the shape of the signal. First, we introduce a measure of the asymmetry of the signal, $\hwhmdiff$, which we define as the difference in width of the trough measured at half its maximal amplitude,
\begin{equation}
    \hwhmdiff = |\nu^{+} - \nu_{\min}| - |\nu^{-} - \nu_{\min}| \label{eq:asymmetry} 
\end{equation}
where $\nu^{+}$ and $\nu^{-}$ refer to the frequencies above and below the extremum at half its maximal amplitude, respectively, and $\nu_{\min}$ is the frequency of the extremum itself. A value of $\hwhmdiff = 0$ indicates a symmetric absorption trough, at least at its half-max point, while positive (negative) values arise when the signal is skewed to higher (lower) frequencies\footnote{We found $\hwhmdiff$ to be a relatively intuitive measure of asymmetry, though there are almost certainly other useful ways to quantify the asymmetry of the signal. We explored several, including the width asymmetry at different amplitudes, slopes at different amplitudes, and the skewness, and found no compelling reason at this time to prefer one over the other.}. Our PopII-only models tend to be only mildly skewed, with $\hwhmdiff \sim 0 \pm 5$ MHz, and a smaller tail out to $\hwhmdiff \lesssim -10$ MHz. The most negatively skewed realizations occur when Wouthuysen-Field coupling is strong at early times but heating occurs late or not at all.

We also quantify the prominence of the ``wings'' of the signal as the ratio of the trough's full-width at half-max\footnote{We prefer the FWHM to the standard deviation at this time since extraction of the latter is likely to be more model-dependent given its dependence on the entire spectrum.} (FWHM) to its depth,
\begin{equation}
    \squash = \frac{\mathrm{FWHM}}{\dTb(\nu_{\min})} \label{eq:squash}
\end{equation}
For PopII-only models, we find that $\squash \sim 0.26 \pm 0.06 \ \slopeunits$, i.e., they are typically $\sim 3-5$ times deeper than they are wide.

In the right panel of Figure \ref{fig:popII_soa}, we also show the mean slope of the signal at frequencies between the first two extrema, $\meanslopelo$, and at frequencies between the absorption minimum and emission maximum, $\meanslopehi$. These quantities were also recently studied in the semi-numeric models of \citet{Cohen2017}, who found a broader range of possibilities, with $-8 \lesssim \meanslopelo / \slopeunits \lesssim -1$ and $1 \lesssim \meanslopehi / \slopeunits \lesssim 6$, in most cases. For now, we simply note that our PopII models span a narrower range of values, which seems to indicate the effects of calibrating to high-$z$ LF measurements. We defer a more detailed comparison of our model sets to future work.

\subsubsection{Expanding the Set of PopIII Models} \label{sec:popIII_gs_grids}
To facilitate comparison with the large set of PopII-only models, we generated a broad set of PopIII models first in a two-dimensional grid over $\Tc$ and $\Ec$, including values as small as the lifetime of very massive stars ($\Tc \sim 2.5$ Myr) and the binding energies of typical supernovae ($\Ec \sim 10^{51}$ erg), up to a factor of 100 more in each dimension. In subsequent sections, we explore also the effects of $\SFRIII$ and $\fXIII$, resulting in a final PopIII model grid with four dimensions.

\begin{figure}
\begin{center}
\includegraphics[width=0.48\textwidth]{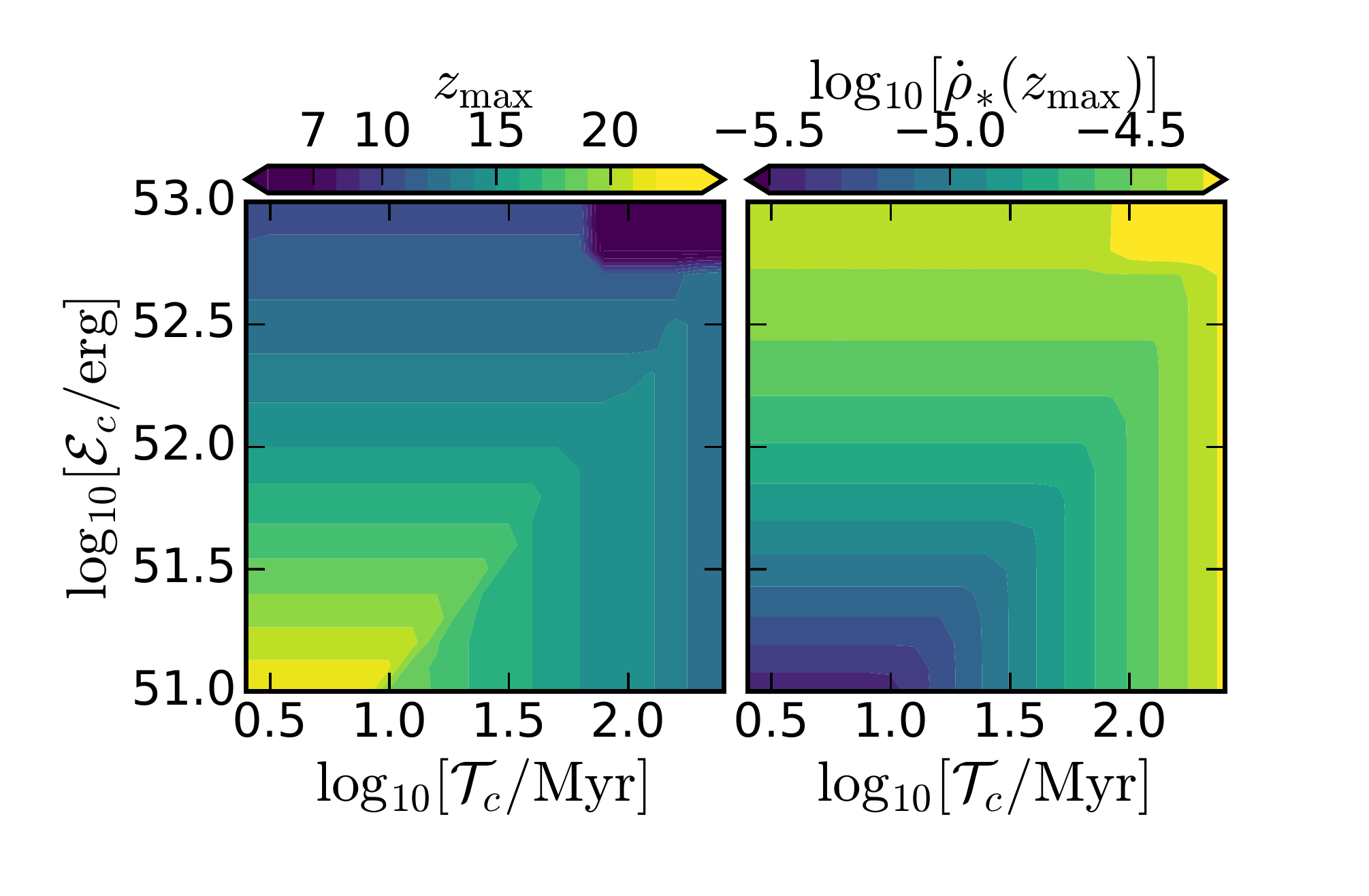}
\caption{Full set of PopIII SFR models for $\SFRIII = 10^{-5} \ \sfrunits$. Each panel shows a 2-D slice of our PopIII star formation model, $\Tc$ vs. $\Ec$, colour-coded by different metrics of the PopIII SFRD. PopIII SFRDs that peak early ($z_{\max} \gtrsim 20$) reside in the lower-left portion of our parameter space (left panel), while peak PopIII SFRD increases monotonically as $\Tc$ and $\Ec$ grow (right panel).}
\label{fig:popIII_2d_grid}
\end{center}
\end{figure}

In Figure \ref{fig:popIII_2d_grid}, we show the bulk properties of the SFRD  in this expanded grid of models, including the redshift at which the PopIII SFRD reaches its maximum, $z_{\max}$, and the SFRD at that redshift, $\dot{\rho}_{\ast,\max}$.

The most salient features of this plot are the anti-correlation between $z_{\max}$ and both $\Tc$ and $\Ec$ (left), and the factor of $\sim 10$ range in peak SFRDs (right). $\Tc$ and $\Ec$ are largely independent, apart from a very mild correlation in the timing of the peak SFRD at small values of each parameter. 

\subsubsection{PopII vs. PopIII} \label{sec:distinguish}
Now, with large sets of PopII-only models and models with PopII and PopIII sources, we can look for distinguishing characteristics of the different model families.

\begin{figure*}
\begin{center}
\includegraphics[width=0.98\textwidth]{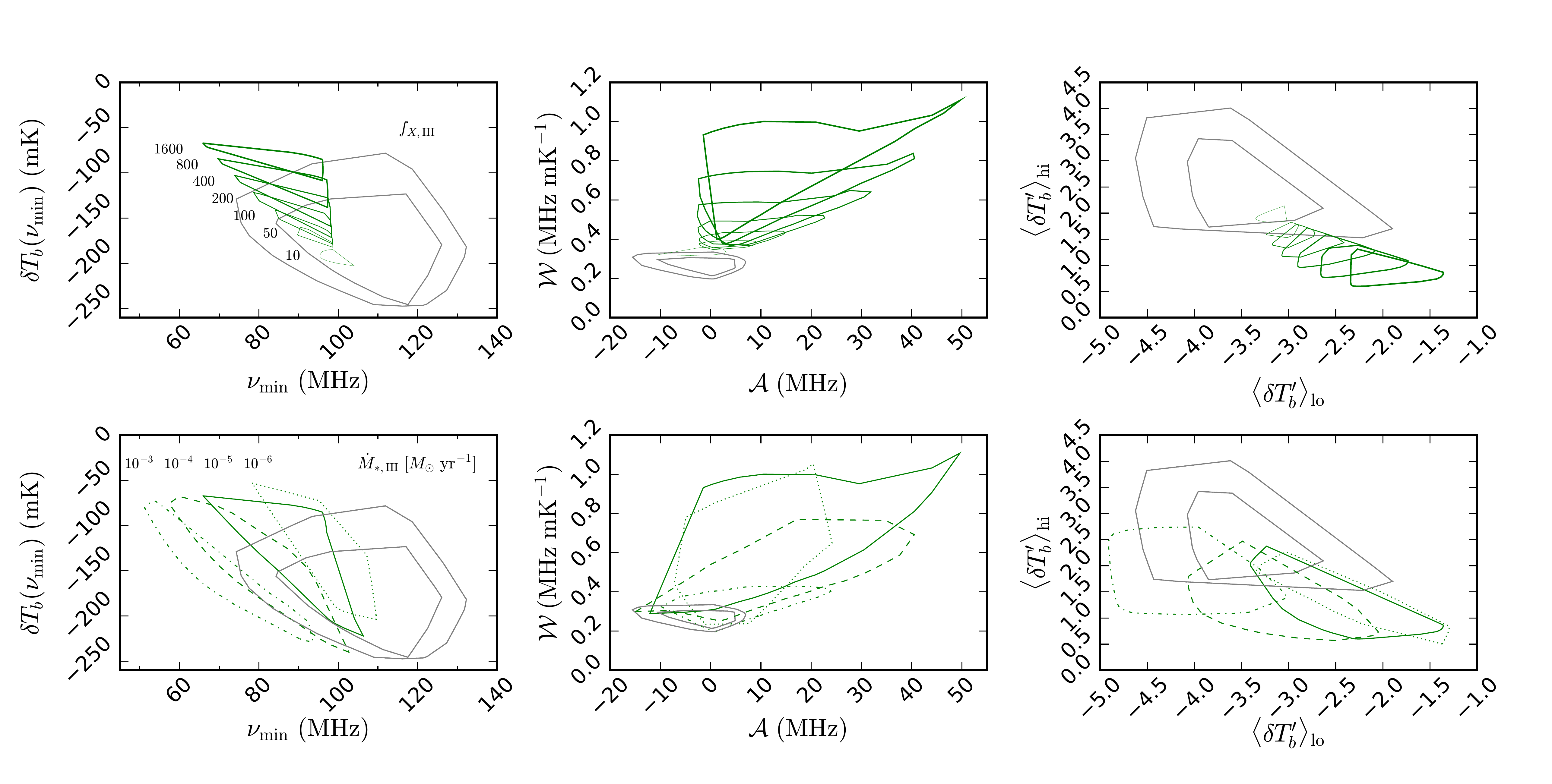}
\caption{Comparison of PopII and PopIII models in three diagnostic spaces, including the position of the absorption trough (left), the prominence of its wings and its asymmetry (middle), and the mean slopes at frequencies above and below the extremum (right), as in Figure \ref{fig:popII_soa}. Black contours enclose sets of PopII models (identical to those in Figure \ref{fig:popII_soa}), while the green polygons are slices through our PopIII model grid, first assuming $\SFRIII=10^{-5} \ \sfrunits$ and select $f_{X,\three}$ values (top row), and then for a set of $\SFRIII$ values having `marginalized' over all $\fXIII$ (bottom row). Measurements falling in regions of overlap between the green and black contours would have no clear evidence of PopIII, while measurements falling only within the green contours would be suggestive of PopIII.}
\label{fig:polygons_II_III}
\end{center}
\end{figure*}

In Figure \ref{fig:polygons_II_III}, we take slices through our set of PopIII models at fixed $\fXIII$ (top) and $\SFRIII$ (bottom), and compare to the entire set of PopII-only models (black polygons, identical to those in Figure \ref{fig:popII_soa}). For clarity, we do not attempt to colour-code each point in the PopIII parameter space by, e.g., $z_{\max}$ or $\rhostardot(z_{\max})$, but simply draw a boundary around the entire set of models in $(\Tc,\Ec)$ space, as we did for the PopII models. As a result, the shapes in Figure \ref{fig:polygons_II_III} simply represent the square ($\Tc,\Ec$) parameter space reprojected into three new planes.

While it is clear that most PopIII models are indistinguishable from PopII-only models based on the trough's position alone (except if $\fXIII \gtrsim 1600$; upper left panel), there is far less overlap between the PopII and PopIII models if one focuses on the shape metrics defined by Equations \ref{eq:asymmetry} and \ref{eq:squash} (middle column), or the mean slope of the signal on either side of the trough (right column). This suggests that detailed measurements of the shape of the global 21-cm absorption trough could reveal the presence of PopIII star formation, even if the position of the trough is consistent with PopII-only models.

For example, consider a scenario in which $\fXIII = 100$. Based on even a precise measurement of the trough, we would be unable to discern the presence of PopIII sources, as the third smallest green polygon sits entirely within the polygon representing even the refined set of PopII-only models. However, in ($\mathcal{A}$, $\mathcal{W}$) space (middle panel of Figure \ref{fig:polygons_II_III}), a measurement of $\mathcal{W} \gtrsim 0.3 \ \slopeunits$ and $\mathcal{A} \gtrsim 7$ MHz would be strong evidence for PopIII.

Unfortunately, moving beyond a simple null test to constrain the parameters of our PopIII model will very difficult. This is in part due to uncertainties in the PopII component of the model, but also due to degeneracies between the PopIII model parameters.

From Figure \ref{fig:polygons_II_III}, we can see some of these degeneracies already. The size of each green polygon in the top row of Figure \ref{fig:polygons_II_III} is set by the range of $\Tc$ and $\Ec$ values we explore, and thus give some indication of the signal's sensitivity to the details of the PopIII SFRD at fixed $\fXIII$ and $\SFRIII$. As a reminder, with $\SFRIII=10^{-5} \ \sfrunits$, the peak SFRD in our $(\Tc, \Ec)$ grid varies by only a factor of $\sim 10$, while $6 \lesssim z_{\max} \lesssim 25$. As a result, measurements with error bars of order the size of the green polygons are probing the peak PopIII SFRD at the order of magnitude level. 

However, there is also significant overlap between the polygons in the right two columns of Figure \ref{fig:polygons_II_III}, which indicates the degeneracy between the PopIII SFRD and $\fXIII$. Though the green polygons in the upper left panel of Figure \ref{fig:polygons_II_III} are distinct, one must remember that each adopts the same value for $\fXII$. As a result, differentiating the effects of $\fXIII$ from the detailed shape of the PopIII SFRD will require additional constraints.

The results presented in this section so far have adopted $\SFRIII=10^{-5} \ \sfrunits$. Boosting this parameter will of course affect the PopIII SFRD, making the signature of PopIII stronger, at least in the $\dTb(\nu_{\min})$ plane, which most intuitively responds to amplification of the SFRD. In the bottom row of Figure \ref{fig:polygons_II_III}, we explore the effects of $\SFRIII$. For $\SFRIII=10^{-4} \ \sfrunits$ (dashed), most PopIII models now have troughs that are distinct from the refined set of PopII models (i.e., the inner polygon), while $\SFRIII=10^{-3} \ \sfrunits$ (dotted) makes PopIII distinct even for the more conservative, complete set of PopII-only models. Alternatively, increasing $\SFRIII$ reduces the signature in $(\hwhmdiff,\squash)$ space (lower middle panel) and in the derivative of the signal (lower right panel), though each metric retains at least some sensitivity to PopIII. Note that these scenarios may be quite extreme, as they require the formation of tens or hundreds of $\sim 100 \ \Msun$ stars per PopIII halo every 10 Myr (for further discussion, see \S\ref{sec:dramatic_departures}). 

\subsection{Side Effects of Persistent PopIII Star Formation} \label{sec:side_effects}
If PopIII stars keep forming until late times, they may measurably influence -- and perhaps violate -- pre-existing constraints. To address this concern, in this section we focus on whether global 21-cm spectra with strong signatures of PopIII cause tension with current limits on the reionization history and the $z=0$ cosmic X-ray background intensity.

So far, in order to isolate the effects of heating and ionization, we have assumed that the escape fraction of Lyman-continuum photons is zero for PopIII halos. However, some simulations predict that LyC escape fractions can be a strong function of halo mass \citep[though this is likely sensitive to resolution;][]{Ma2015}, rendering PopIII halos a potentially important source population to consider for reionization \citep[e.g.,][]{Wise2009}. However, star formation rates in minihalos may be low enough to counteract large escape fractions, confining their effects to the earliest stages of reionization \citep{Xu2016}, thus rendering them unimportant to the bulk of the process \citep{Kimm2017}. 

\begin{figure}
\begin{center}
\includegraphics[width=0.49\textwidth]{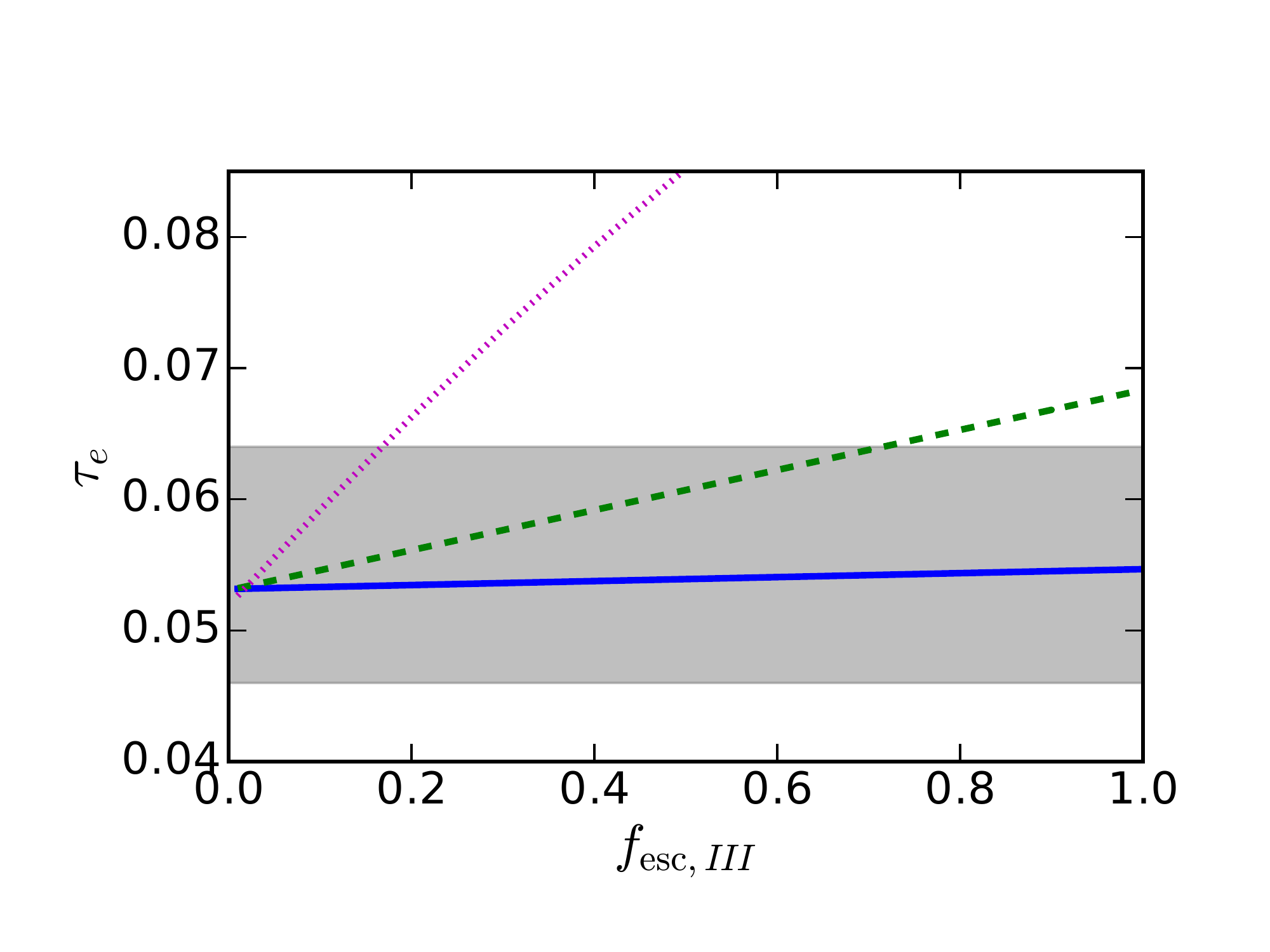}
\caption{CMB optical depth as a function of PopIII LyC escape fraction. Solid, dashed, and dotted curves assume increasingly persistent PopIII star formation, from $\Tc=2.5$ Myr and $\Ec=10^{51}$ erg (solid), to $\Tc=250$ Myr and $\Ec=10^{53}$ erg, with $\SFRIII=10^{-5} \ \sfrunits$. For reference, the exact set of corresponding SFHs and global 21-cm spectra are shown in the upper-right panels of Figure \ref{fig:popIII_sfrd} and \ref{fig:popIII_gs_XRIII}. 68\% confidence interval from \citet{Planck2016} is shown in gray.}
\label{fig:popIII_tau}
\end{center}
\end{figure}

Adopting our default model, that assumes massive PopIII stars that form at a rate $\SFRIII=10^{-5} \ \sfrunits$, we investigate the impact of increasing $\fescIII$ on the CMB optical depth, $\tau_e$, in Figure \ref{fig:popIII_tau}. For our default PopII-only model with $\fescII=0.1$, the addition of PopIII stars that form only for a short time in high-$z$ halos (solid blue) affects $\tau_e$ negligibly, even if $\fescIII=1$. The corresponding SFRD of this model is identical to the solid blue line in the upper right panel of Figure \ref{fig:popIII_sfrd}. If we instead assume $\Tc=250$ Myr and $\Ec=10^{53}$ erg, PopIII star formation is much more plentiful at late times, driving $\tau_e$ outside the preferred \textit{Planck} 2-$\sigma$ range if $\fescIII \gtrsim 0.4$ (dotted magenta). In between, e.g., if $\Tc=25$ Myr and $\Ec=10^{52}$ erg, $\fescIII = 1$ is still allowed at the $2\sigma$ level.

These results are again driven by our adoption of parameters consistent with massive $\sim 100 \ \Msun$ stars, which emit $\Nion \sim 10^5$ ionizing photons per stellar baryon. If the PopIII IMF is more normal, and $\Nion$ is reduced to values of 5-$10 \times 10^3$, as is appropriate even for metal-poor PopII sources, then PopIII sources will have a negligible impact on $\tau_e$ unless they form at rates substantially higher than $\SFRIII=10^{-5} \ \sfrunits$. Though difficult to compare closely, this is roughly in agreement with \citet{Visbal2015} and \citet{Sun2016}.
 
Next, because PopIII sources are most noticeable in the global 21-cm signal when their X-ray production efficiencies are high $\fXIII \gg 1$, we investigate whether such scenarios violate constraints on the $z=0$ X-ray background intensity. To do so, we extract the cosmic X-ray background spectrum at $z=6$ (when our calculations terminate) and evolve the background to $z=0$ assuming no attenuation by the IGM. This is a reasonable assumption since the $z=0$ soft X-ray background (typically defined as the 0.5-2 keV band) probes rest-frame photon energies of 3.5-14 keV at $z=6$, which will be optically thin even to dense absorbers in the IGM until relatively late times.

\begin{figure}
\begin{center}
\includegraphics[width=0.48\textwidth]{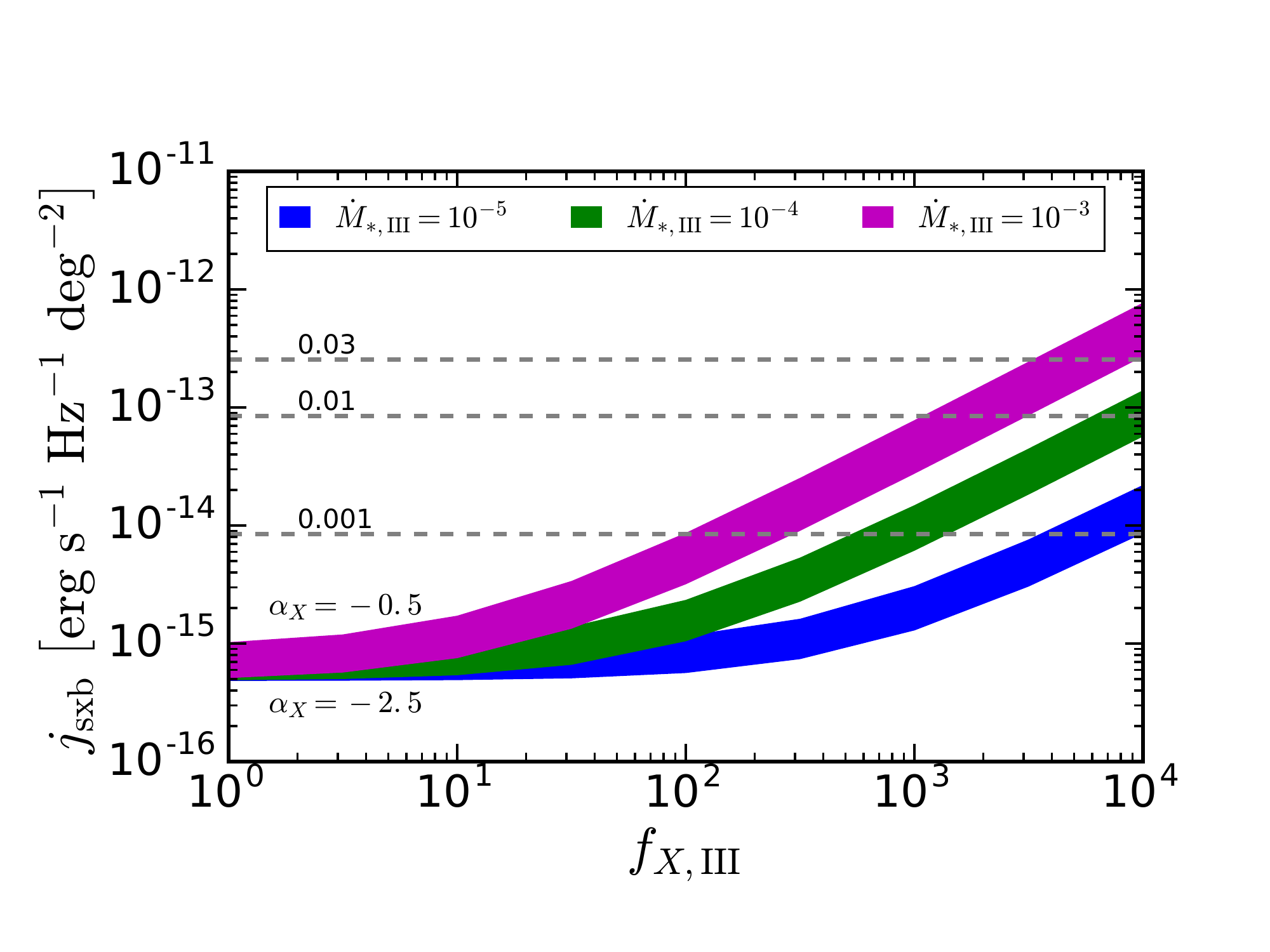}
\caption{Cosmic X-ray background intensity at $z=0$ generated by our models as a function of $\fXIII$ ($x$-axes) and $\SFRIII$ (coloured bands). The width of each band corresponds to a difference in the assumed high energy power-law index, $\alpha_X$, as indicated in the lower left portion of the plot. We assume our default PopII model with $\fsc=0.1$ and $\alpha_X=-2.5$, which sets the floor at $10^{-15} \ \CXRBunits$. Dashed lines indicate fluxes at a fixed fraction of the \citet{Cappelluti2012} measurement of the 0.5-2 keV CXRB flux at $z=0$. The \citet{Cappelluti2012} models find that the unresolved fraction is $\lesssim 3$\%, meaning our modeled PopIII sources are only in tension with the CXRB if $\fXIII \gtrsim 10^3$ and $\SFRIII \gtrsim 10^{-3} \ \sfrunits$.}
\label{fig:popIII_cxrb}
\end{center}
\end{figure}

In our default case of a pure MCD X-ray spectrum, there is relatively little emission at high energies $\gtrsim 10$ keV, which will artificially bias our predictions for the $z=0$ CXRB low. To explore a more realistic case, we instead use the SIMPL model \citep{Steiner2009} as our input X-ray spectrum, and assume that a fraction $f_{\mathrm{sc}}=0.1$ of accretion disk photons are up-scattered to a high energy tail, in which emission follows a power-law of index $\alpha_X$. The introduction of these modifications has a very minor effect on the thermal history. As a result, one can compare models for the global signal and CXRB at fixed $\fXIII$ at the level of $\sim 10$ mK and $\sim$ few MHz in the location of the absorption trough \citep[see Figures 4 and 5 of][]{Mirocha2014}.

In Figure \ref{fig:popIII_cxrb}, we find that PopIII sources alone provide an essentially negligible contribution to the $z=0$ CXRB, unless $\fXIII \gtrsim 10^3$ \textit{and} $\SFRIII \gg 10^{-5} \ \sfrunits$. This is a corner of parameter space that is perhaps unreasonable theoretically, as it requires extremely efficient PopIII star formation, which can likely only occur if the IMF is fairly normal. But, if the IMF is normal, values of $\fXIII \sim 10^3$ are much more difficult to explain. We will discuss this further in Section \ref{sec:dramatic_departures}.

Notice that all curves approach $j_{\mathrm{sxb}} \sim 10^{-15} \ \CXRBunits$ when $\fXIII$ is small. This is the CXRB intensity produced in our default model from PopII sources only (i.e., roughly 0.01\% of the unresolved background). If we have underestimated the contribution of PopII galaxies to the present day CXRB by a factor of 10 or 100, this will raise the floor in $j_{\mathrm{sxb}}$ in Figure \ref{fig:popIII_cxrb}, leaving less room for PopIII sources. Any further observational reduction in the unresolved fraction would similarly limit $\fXIII$ to smaller values. If that reduction were due to low redshift sources only, it could help improve lower limits on the strength of the global 21-cm absorption signal \citep{Fialkov2017}.

To summarize, it appears that strong signatures of PopIII stars in the global 21-cm signal can arise without violating pre-existing limits on $\tau_e$ or $j_{\mathrm{sxb}}$. This could be subject to revision if either the UV or X-ray backgrounds generated by PopII sources are much stronger (and/or harder) than is assumed in our default model, or if more of $\tau_e$ and the unresolved fraction can be attributed to normal star-forming galaxies. 

\section{Discussion} \label{sec:discussion}
We have found that massive PopIII stars can give rise to broad, asymmetric wings in the global 21-cm signal that do not arise in PopII-only models. While such sources also affect the position of the absorption minimum,  uncertainties in the properties of PopII sources could prevent an unambiguous detection of PopIII based on the trough position alone. In this section, we focus on the observational implications of these findings, as well as so-far-neglected elements of the theoretical modeling that, if treated explicitly, could lead to ambiguity in the interpretation of the signal's shape.

\subsection{Implications for Global 21-cm Experiments} \label{sec:implications}
By construction, our results agree with those of \citet{Mirocha2017} in that our most `vanilla' models for the global 21-cm signal predict a strong absorption trough near $\sim 100$ MHz. Though adding a PopIII component to the model can dramatically suppress the amplitude of the absorption signal, its location in frequency remains largely unaffected unless PopIII star formation is very efficient ($\gtrsim 10^4 \Msun$ per event). The critical finding in this work is that shallow absorption troughs driven by PopIII sources are \textit{not} accompanied by strong emission features at high frequencies, which leads to an asymmetry in the width of the trough. This is in contrast to scenarios in which PopII sources dominate the X-ray background at all epochs, in which case shallow troughs are accompanied by strong emission signals since the PopII SFRD is monotonically rising as reionization progresses, and the X-ray emission is assumed to trace star formation. In these cases, the global 21-cm signal remains relatively symmetrical ($|\hwhmdiff| \lesssim 5-10$ MHz).

While most forecasting work has focused on the ability of various experimental setups to constrain the locations of extrema in the global 21-cm signal \citep[e.g.,][]{Pritchard2010a,Harker2012,Liu2013,Harker2016}, and perhaps the width of the absorption trough \citep{Presley2015,Bernardi2015}, to our knowledge there has been no investigation of the information content of the trough's symmetry or prospects for its characterization. We suspect that recovery of shape information will in general be more difficult than recovery of the `turning points,' especially for the PopIII models, whose shallow late-time gradients will increase their resemblance to the galactic foreground. 

Though the asymmetry has not been quantified in previous works, \citet{Cohen2017} investigated the mean slope between the minimum and maximum of the signal in a set of of semi-numeric models and found $1 \gtrsim \meanslope / (\mathrm{mK} \ \mathrm{MHz}^{-1}) \gtrsim 6$. Our PopII-only models span a narrower range, $1.5 \gtrsim \meanslope \gtrsim 3$, while our PopIII models exhibit $\meanslope \lesssim 1.5 \ \mathrm{mK} \ \mathrm{MHz}^{-1}$ (see Figure \ref{fig:popII_soa}). It is challenging to make a direct comparison given the difference in methods, but it is at least encouraging to see that the \citet{Cohen2017} models rarely produce realizations with $\meanslope \lesssim 2 \ \mathrm{mK} \ \mathrm{MHz}^{-1}$, since there was no attempt to treat PopIII sources in detail in their work.

Observational efforts have recently commenced attempts to rule out `cold reionization' scenarios \citep{Singh2017,Monsalve2017}, which have the potential to produce the strongest absorption signals. In our framework, they should be treated as the null hypothesis, as one arrives at strong late-peaking absorption signals when constructing minimal models of high-$z$ galaxies. In some sense, simply ruling these models out may provide the first evidence of `new' source populations at high-$z$. However, such claims require high confidence in the PopII SFRD (and its extrapolation) and $L_X$-SFR relation of PopII galaxies at high-$z$.

Future data analysis pipelines could call our models directly and attempt to fit for the parameters of PopII and PopIII sources. Even for simple models like ours, this can be fairly expensive, and may also be somewhat restricting since our models cannot fit an arbitrary signal. In practice, it may be more economical to use a flexible parametric form for the signal that is capable of generating a wide variety of realizations, and simply compute $\hwhmdiff$ and $\squash$ in post-processing. A crude form of model selection could then be applied simply by asking whether or not the best-fitting value is consistent with PopII-only models (using, e.g., our contours in Figures \ref{fig:popII_soa}-\ref{fig:polygons_II_III}), or if an additional component is required. 

Should the calibration of the PopII component of the model mature in the coming years, precise measurements may be able to move beyond a simple null test and attempt to constrain the parameters of the PopIII model. We defer a detailed forecast of this possibility to future work.

Finally, before concluding, we reconsider the possibility that the true global 21-cm signal qualitatively differs from the \citet{Mirocha2017} predictions of a strong, high-frequency trough (\S\ref{sec:dramatic_departures}), and comment on effects that we have neglected that could potentially induce asymmetry in the global 21-cm signal, and thus provide a source of confusion in future studies (\S\ref{sec:confusion}).

\subsection{Implications of Dramatic Departures from $\sim 100$ MHz Troughs} \label{sec:dramatic_departures}
In \citet{Mirocha2017}, we suggested that observational rejections of our models would most likely indicate the need for ``new'' source populations -- i.e., those deviating strongly from the inferred (or extrapolated) properties of high-$z$ galaxies. Given that we explicitly neglected star-formation in minihalos, such an outcome would not spell disaster for galaxy formation models, but only serve to emphasize the importance of very low-mass objects in the early Universe. Having since added low-mass halos in a physically-motivated way, our predictions for the amplitude of the absorption trough cover a broader range, though it is still difficult to achieve low frequency absorption troughs. What would the implications of a $\nu_{\min} \ll 100$ MHz absorption trough mean \textit{now}?

In order to produce realizations with early troughs, we need to revise the model in one or more of the following ways:
\begin{enumerate}
    \item Allow massive PopIII stars to form in large numbers (tens or hundreds per 10 Myr; our $\SFRIII > 10^{-5} \ \sfrunits$ cases), as explored in Figures \ref{fig:popIII_sfrd} (bottom two rows) and \ref{fig:polygons_II_III}.
    \item Turn off global LW feedback, which keeps $\Mmin$ at low levels and boosts the abundance of PopIII star-forming halos as a result.
    \item Introduce a floor in the PopII star formation efficiency so that when PopIII halos transition to PopII, their SFE is much larger than extrapolation of our default double power-law SFE would predict, leading to a stronger UV background capable of triggering Wouthuysen-Field coupling earlier.
    \item A strong, but globally short-lived epoch of PopIII star formation, that can induce a second trough at low frequencies $\nu \sim 50-60$ MHz. Though the typical trough would still occur at higher frequencies, the `dramatic' designation still seems appropriate.
\end{enumerate}

Let us entertain each of these possibilities in turn.

First, the formation of massive $\sim 100 \ \Msun$ PopIII stars in large numbers ($\SFRIII > 10^{-5} \ \sfrunits$) may be physically unrealistic. For example, it seems unlikely that the small halos could recover quickly from such dramatic bursts of star formation, which are almost surely accompanied by many supernovae\footnote{Unless most or all PopIII stars undergo direct collapse to a black hole.}. Long recovery times could thus counteract intense star formation episodes, keeping $\SFRIII$ at low, relatively constant, levels.

One way around this is to assume that the PopIII IMF is at least somewhat normal, in which case the effects of supernovae feedback on the galactic gas supply could be reduced and recovery times might remain short, even in the face of $\sim 10^3$ or $10^4 \ \Msun$ star-forming events. However, changing the IMF in this way also reduces the strength of the UV background by a factor of $\sim 5-10$, since low-mass stars produce UV photons less efficiently than very massive stars. As a result, the impact of such a population on the low-frequency part of the global 21-cm signal will also be reduced. In order to shift the absorption trough to low frequencies (as in the bottom two rows of Figure \ref{fig:popIII_sfrd}), one would need to enhance $\SFRIII$ by an extra factor of $\sim 5-10$ more than the values quoted along the edge of Figure \ref{fig:popIII_sfrd}.

Furthermore, if the PopIII IMF is normal, it might be difficult to justify $\fXIII \gg 10$ following the arguments of Equations \ref{eq:LHMXB}-\ref{eq:NHMXB}, further reducing the impact of such sources on the global 21-cm signal unless $\SFRIII$ is very large ($\gtrsim 10^{-3}$). These arguments support the PopII-only predictions of \citet{Mirocha2017}, which favor an absorption trough at relatively high frequencies, $\nu_{\min} \sim 100$ MHz. Clearly, as we have shown, PopIII stars can modify the amplitude of the trough, though substantially altering its frequency -- and recovering models with $\sim 70$ MHz troughs  \citep[as have been common in recent years;][]{Furlanetto2006,Pritchard2010a,Mirocha2015} -- may require unreasonably efficient PopIII star formation. Fortunately, the asymmetry (though perhaps subtle), appears to emerge even if both the efficiency of PopIII star formation and  $\fXIII$ are relatively small. 

PopIII IMF effects aside, low-frequency troughs may also arise if global LW feedback is weaker than expected, or non-existent. One way to reduce (or eliminate) feedback is to explicitly treat the interplay between the UV and X-ray backgrounds. Because X-rays are able to travel through dense proto-stellar clouds, they can boost the free electron fraction upon absorption and thus catalyze the formation rate of $H_2$ and fend off the destruction of $H_2$ by LW photons \citep{Machacek2003,Glover2016,Ricotti2016}. The resultant decline in $\MminIII$ permits PopIII star formation in halos of lower mass than the usual relation between $\JLW$ and $\MminIII$ would suggest, enhancing the PopIII SFRD, and thus all radiation backgrounds generated by PopIII sources.

Next, though there is some indication that the slope of the PopII star formation efficiency may become more shallow at low mass \citep{Mason2015}, we can think of no obvious reason why PopII halos would have a minimum star formation efficiency. Continued efforts to measure the galaxy luminosity function at higher redshifts and fainter magnitudes should help explore this possibility in the coming years.

Finally, a single $\sim 100$ MHz trough could be accompanied by another trough at low frequencies, which in our model occurs only in situations of very efficient, though globally brief (i.e., $z \gtrsim 20$ only), PopIII star formation. This only happens if the transition to PopII occurs at a fixed halo binding energy, though this is somewhat unphysical, as passage through a critical binding energy does not influence the cooling properties of gas in a halo. However, it is possible for halos to skip the \textit{massive} PopIII phase via a different mechanism: arrival at the atomic-cooling threshold before forming their first stars. In this case, cooling is expected to be efficient \citep{Susa1998,Nakamura2002,Oh2002}, and lead to a stellar IMF less top-heavy than that which arises in molecular-cooling halos, though the stars are still technically PopIII as they have zero metallicity. This has led to a distinction between PopIII.1 stars (born in molecular-cooling halos) and PopIII.2 stars (born in atomic-cooling halos, or from previously ionized gas).

\citet{Mebane2017} showed that many halos can skip the PopIII phase if the minimum mass rises rapidly at early times, which is easiest to achieve if the PopII SFRD is high at early times (since PopII halos do not ``feel'' LW feedback). As a result, this could be related to our previous point, regarding the potential for a shallower PopII SFE, which would boost the PopII SFRD at early times. Short of introducing these kinds of stars as a new source population in our model, it is most sensible to count this as PopII star formation, since the LW and X-ray yields would be less extreme for these halos than the molecular-cooling halos in the PopIII phase. As a result, their impact on the global 21-cm signal (and PISN supernova rates, for that matter), will be fairly minimal.

While drawing a distinction between the PopIII.1 and PopIII.2 sources may have important consequences for the global 21-cm signal, we note that the predicted PopIII SFRD in this scenario in \citet{Mebane2017} is small ($\sim 10^{-6} \ \sfrdunits$ at peak), compared to the $\gtrsim 10^{-5} \ \sfrdunits$ realizations that led to a double trough in this study (see Figure \ref{fig:popIII_sfrd}). Furthermore, because of the efficient PopII star formation, the $\Lya$ background would be strong and rise monotonically with time, thus preventing the emergence of a double trough. Rejection of the double trough scenario observationally may thus be capable of ruling out a very specific -- though unlikely -- set of circumstances in high-$z$ star-forming galaxies.

\subsection{Potential for Confusion} \label{sec:confusion}
In the previous section, we focused on effects that could push the absorption minimum to low frequencies, beyond what we can comfortably accomplish within the confines our LF-calibrated models. In this section, we focus instead on the possibility of more subtle effects that we have yet to include that may weaken our central claims.

To phrase our results very conservatively, broad asymmetric wings in the global 21-cm signal are indicative of a slowly rising (or flat) heating rate density at high-$z$. It seems natural to associate this with PopIII star-forming halos, since numerous studies have predicted shallow PopIII SFRDs, and thus luminosity densities at all wavelengths. But, there could be other mechanisms at work, either on the PopII or PopIII side of the model, that could complicate such expectations.

For example, a source population that only emits X-rays at early times, or produces X-rays at a very gradual rate, could mimic the signature of our PopIII model. Dark matter annihilation may provide such a source, as the heating rates in most models evolve very gradually with redshift \citep[e.g.,][]{Valdes2013,LopezHonorez2016}. We plan to address this potential source of confusion in future work.

We have also neglected emission from super-massive black holes, whose growth rate may not mirror the growth rates of galaxies at early times, and could thus generate radiation backgrounds that are decoupled from the cosmic SFRD. However, heating rates from rapidly growing stellar remnants and direct collapse BHs are both still expected to be quite steep in redshift \citep{Tanaka2016}. If such predictions hold, neglect of such sources may bias inferred values of the SFRD and $f_X$ values for PopII and PopIII sources, but would be unlikely to introduce much asymmetry to the signal.

More mundane source of gradual heating, i.e., those which originate in `normal' galaxies, are perhaps more difficult to imagine since they require that the galaxy population become a less efficient producer of X-rays \textit{per unit star formation} on timescales shorter than $\sim 1$ Gyr. Only then can the rapid rise in the PopII SFRD be offset to produce a gradual heating rate. 

For example, because X-ray production likely becomes less efficient with increased metallicity, which ought to rise monotonically in galaxies with time, one might think this a reasonable way to generate gradual heating without PopIII sources. However, the PopII SFRD is expected to rise by $\sim 3$ orders of magnitude over $6 \lesssim z \lesssim 20$, whereas $L_X$/SFR only appears to decline by a factor of $\sim 10$ between $Z=10^{-3}$ and solar metallicity \citep{Brorby2016}. As a result, rapid evolution in the properties of X-ray binaries is unlikely to be able to counteract the rapid rise in the SFRD.

\section{Conclusions} \label{sec:conclusions}
Our main results can be summarized as follows:
\begin{enumerate}
    \item PopIII stars can affect the global 21-cm signal through both their UV and X-ray emissions. Because the PopIII SFRD flattens, and sometimes declines before reionization is complete, boosts in the LW and X-ray backgrounds (relative to the PopII-only case) occur mostly at the highest redshifts, and thus mostly affect the low frequency portion of the global 21-cm signal. This leads to broad absorption troughs skewed toward high frequencies.
    \item We explore two ways to quantify this modulation of the signal: the asymmetry, $\mathcal{A}$, and the prominence of the wings of the signal, $\mathcal{W}$, both of which seem to clearly identify the presence of PopIII sources, even in cases where the trough position is consistent with a PopII-only model.
    \item These models reinforce the \citet{Mirocha2017} predictions, since PopIII star formation must be extremely efficient to drive the signal to frequencies below $\sim 90$ MHz. Fortunately, the signature of PopIII -- albeit subtle -- appears even in the likely event that their overall SFRD is small.
\end{enumerate}

This work was supported by the National Science Foundation through awards AST-1440343 and 1636646, by NASA through award NNX15AK80G, and was completed as part of the University of California Cosmic Dawn Initiative. In addition, this work was directly supported by the NASA Solar System Exploration Research Virtual Institute cooperative agreement number 80ARC017M0006. We acknowledge support from the University of California Office of the President Multicampus Research Programs and Initiatives through award MR-15-328388. K.S. gratefully acknowledges support from NSF Grant PHY-1460055 through the UCLA Physics \& Astronomy REU program. D.T. acknowledges support from the University of California's Leadership Excellence through Advanced DegreeS (UC LEADS) program. This work used computational and storage services associated with the Hoffman2 Shared Cluster provided by UCLA Institute for Digital Research and Education's Research Technology Group.

\textit{Software:} \textsc{python}, and packages \textsc{matplotlib}, \textsc{numpy}, \textsc{scipy}, \textsc{descartes}, and \textsc{shapely}.

\bibliography{references}
\bibliographystyle{mn2e_short}

\end{document}